\documentclass[a4paper,superscriptaddress,preprint,prb]{revtex4}
\pdfoutput=1 

\usepackage[fleqn]{amsmath}
\usepackage{amsfonts,amssymb,bm}
\usepackage[pdftex,bookmarksnumbered]{hyperref}
\usepackage{graphicx}
\usepackage{multirow}

\hypersetup{pdfauthor={Juan J. Cerda, V. Ballenegger, O. Lenz and
    C. Holm}, pdftitle={P3M algorithm for dipolar interactions}}
\newcommand{\todo}[1]{}

\newcommand{\es}{\textsf{ESPResSo}}             
 
\newcommand{\dotprod}{\cdot}
\newcommand{\erf}{\mathrm{erf}}
\newcommand{\erfc}{\mathrm{erfc}}
\newcommand{\self}{\mathrm{(self)}}
\newcommand{\surf}{\mathrm{(surf)}}
\newcommand{\corr}{\mathrm{(corr)}}

\newcommand{\pppm}{\mathrm{P3M}}
\newcommand{\PPPM}{$\mathrm{P^3M}$}
\newcommand{\exact}{\mathrm{(ex)}}
\newcommand{\INT}{\mathrm{int}}
\newcommand{\opt}{\mathrm{opt}}
\newcommand{\rcut}{r_{\mathrm{cut}}}

\newcommand{\FFT}{\mathrm{FFT}}
\newcommand{\FFTs}[1]{\widetilde{#1}}

\newcommand{\FTnp}[1]{\breve{#1}}
\newcommand{\FSp}[1]{\hat{#1}}
\newcommand{\FSpletters}{\mathrm{FT}}

\newcommand{\dd}{\mathrm{d}}
\newcommand{\bracket}[1]{\left\langle #1 \right\rangle}

\bmdefine\bmu{\mu}
\bmdefine\btau{\tau}
\bmdefine\bmeta{k}
\bmdefine\brho{\rho}
\bmdefine\bn{n}
\bmdefine\bmm{m}
\bmdefine\bk{k}
\bmdefine\br{r}
\bmdefine\bx{x}
\bmdefine\bxi{\xi}
\bmdefine\bT{T}
\bmdefine\bF{F}

\bmdefine\bkprima{k^{\prime}}
\bmdefine\bkone{k}   

\newcommand{\SUMOK}{\sum_{\bk \in \FFTs{\mathbb{K}}^3  \atop \bk \ne 0}}

\newcommand{\SUMOKone}{\sum_{\bkone \in \FFTs{\mathbb{K}}^3  \atop \bkone \ne 0}}
\newcommand{\SUMOKprima}{\sum_{\bkprima \in \FFTs{\mathbb{K}}^3 \atop \bkprima \ne 0 }}
\newcommand{\SINGLE}{(\br, \hat{\bmu})}
\newcommand{\NOSINGLE}{(\br, \bmu)}
\newcommand{\SUMOKwithkzero}{\sum_{\bk \in \FFTs{\mathbb{K}}^3}}

\newcommand{\ie}{\emph{i.e.}} 
\newcommand{\eg}{\emph{e.g.}}
\newcommand{\mic}{\emph{m.i.c.}}

\newcommand{\madelungself}{{MS}}
\newcommand{\MS}{\mathrm{ms}}

\begin{document}

\title{\PPPM\ algorithm for dipolar interactions}

\author{Juan J. Cerd\`{a} } \email{jcerda@fias.uni-frankfurt.de}
\affiliation{Frankfurt Inst. for Advanced Studies, (FIAS), Goethe -
  Universit\"{a}t,  Ruth-Moufang Str. 1 ,60438, Frankfurt am Main, Germany}
\author{V. Ballenegger} 
\affiliation{Institut UTINAM, Universit\'e de Franche-Comt\'e, CNRS,
  16, route de Gray, 25030 Besan\c{c}on cedex France.}\
\author{O. Lenz}
\affiliation{Frankfurt Inst. for Advanced Studies, (FIAS), Goethe -
  Universit\"{a}t,  Ruth-Moufang Str. 1 ,60438, Frankfurt am Main, Germany}
\affiliation{Max-Planck-Institut f\"{u}r Polymerforschung, Ackermannweg
10, 55128, Mainz,
  Germany}
\author{C. Holm} 
\affiliation{Frankfurt Inst. for Advanced Studies, (FIAS), Goethe -
  Universit\"{a}t,  Ruth-Moufang Str. 1 ,60438, Frankfurt am Main, Germany}
\affiliation{Max-Planck-Institut f\"{u}r Polymerforschung, Ackermannweg
10, 55128, Mainz,
  Germany}

\date{\today}

\begin{abstract}
  An extension to the \PPPM\ algorithm for electrostatic interactions
  is presented, that allows to efficiently compute dipolar
  interactions in periodic boundary conditions.  Theoretical estimates
  for the root-mean square error of the forces, torques and the energy
  are derived.  The applicability of the estimates is tested and
  confirmed in several numerical examples.  A comparison of the
  computational performance of the new algorithm to a standard dipolar
  Ewald summation methods shows a performance crossover from the Ewald
  method to the dipolar \PPPM\ method for as few as 300 dipolar
  particles.  In larger systems, the new algorithm represents a
  substantial improvement in performance with respect to the dipolar
  standard Ewald method.  Finally, a test comparing point-dipole based
  and charged-pair based models shows that point-dipole based models
  exhibit a  better performance than charged-pair based models.
  
\end{abstract}

\maketitle


\section{Introduction}

Dipolar interactions are important in many soft-matter systems ranging
from dispersions of magnetic micro- and nanoparticles (ferrofluids)
and electrorheological fluids to magnetic thin films and water
\cite{odenbach02a,rosensweig85a,berkovsky93a,berkovsky96a,holm05b,weis05a}.
Numerical simulations play a central role in explaining and
unravelling the rich variety of new and unexpected behavior found in
recent theoretical and experimental studies on dipolar systems
\cite{blums97a,odenbach02b}. Especially for systems which possess
point-dipolar interactions such as dipolar model systems used in
analytical theories, or ferrofluids, a numerical algorithm based on
truly dipolar interactions is needed \cite{holm05b,weis05a}. Periodic boundary conditions are
frequently used in these simulations in order to approach bulk systems
within the limits of currently available computers (see ref. \cite{heinz05a} for a
detailed discussion about the adequacy of such methods to describe electrostatic
systems).  If a system of $N$ particles with positions $\left\{ \br_i \right\}_{i=1}^{i=N}$ in a
cubic box of length $L$ that carry point dipoles $\{ \bmu_i
\}_{i=1}^{i=N}$ is considered, then the total electrostatic energy
under periodic boundary conditions is given, in Gaussian units, by
\begin{gather}
\label{direct_sum}
U = \frac{1}{2} 
        \sum_{i=1}^{N} \sum_{j=1}^{N} \sideset{}{^{'}}\sum_{\bn \in \mathbb{Z}^3}
v(\br_{ij}+\bn L, \bmu_i ,\bmu_j) \\	\notag
\end{gather}
where $\br_{ij} = \br_i - \br_j$, and 
\begin{gather}
v(\br_{ij}, \bmu_i, \bmu_j) \equiv
(\bmu_i \dotprod \nabla_{\br_i}) \left(\bmu_j
\dotprod \nabla_{\br_j} \right)
\frac{1}{|\br_{ij}|}  \notag \\
\phantom{v(\br_{ij}, \bmu_i, \bmu_j)}
=  \frac{\bmu_i \dotprod \bmu_j }{|\br_{ij}|^3} - 
  \frac{3  \left( \bmu_i \dotprod \br_{ij}
  \right)  \left( \bmu_j \dotprod \br_{ij}
  \right)}{|\br_{ij}|^5}
\label{v_dip}
\end{gather}
is the dipolar pair interaction for point dipoles.  The innermost sum
runs over all periodic images of the system, identified by the
shifting integer vector $\bn$. The prime in the sum in
eq.~\eqref{direct_sum} indicates that the $i=j$ term must be omitted
for $\bn=0$. Note that the dipolar sum is conditionally converging\cite{deLeeuw80a} and
its precise value depends on the summation order.
In what follows we assume  eq.~\eqref{direct_sum} to be summed over
spherical shells (spherical order of
summation)\cite{deLeeuw80a,deLeeuw80b}.

 The force $\bm{F}_i$, and the
electrostatic field $\bm{E}_i$ acting on a particle $i$ can be
obtained by differentiating the potential energy $U$ with respect to
${\bm r}_i$ and $\bmu_i$ respectively, \ie{},
\begin{gather} 
\label{eq3}
\bm{F}_i = - \nabla_{\br_i} U \\
\label{eq4}
\bm{E}_i = - \nabla_{ \bmu_i} U.
\end{gather} 

In the case of dipoles, these quantities are related via
$\bm{F}(\br)= \nabla_{\br} (\bmu \dotprod \bm{E}(\br))$.  For
point-dipoles, the torque $\btau_i$ acting on particle $i$ can be
related to the electrostatic field at the position of the particle as
\begin{equation}
  \label{torques}
  \btau_i = \bmu_i \times \bm{E}_i 
\end{equation}
 
Performing the direct summation of the interactions
(eq.~\eqref{direct_sum}) is impracticable beyond a few particles due
to the slow convergence of the innermost sum and the quadratic scaling
with the total number $N$ of particles in the outer sums.  However,
algorithms have been proposed to speed up the computation of dipolar
interactions: the dipolar Ewald sum \cite{allen87a,wang01a}, the
dipolar Lekner sum \cite{weis05a}, the (Smooth) Particle-Mesh Ewald
methods: PME and SPME \cite{toukmaji00a}, and Multipole Methods (MM):
Fast-MM, and Cell-MM
\cite{kutteh95a,kutteh95b,stoycheva02a,christiansen93a}.  For a
general overview of these algorithms, see the reviews in refs.
\cite{weis05a,arnold05a}.

Although the Ewald summation is significantly better than direct
summation from a computational point of view, it still exhibits an
unfavourable $\mathcal{O}(N^{3/2})$ scaling with the number of
particles~\cite{perram88}.  By contrast, Multipole methods scale
linearly, but have a large prefactor in the $\mathcal{O}(N)$
scaling. In the case of point charges, Multipole methods have been
found to be superior to mesh methods only for very large systems $N
\geq 100000$ (see discussion in ref.  \cite{petersen95a} and
\cite{sagui99a}).  For systems of moderate size, optimal algorithms
are those that take advantage of the Fast Fourier Transform ($\FFT$)
in order to compute the Fourier contribution to the Ewald sum, which
are commonly known as particle mesh methods: PME, SPME, and
Particle-Particle-Particle Mesh (\PPPM), which is introduced in this
article. These methods are all $\mathcal{O}(N \log N)$, \ie\ they
exhibit a nearly linear scaling with the number of particles.

When computing Coulomb interactions, the \PPPM\
method~\cite{hockney81a} achieves the highest accuracy among the
particle mesh methods, thanks to its use of the optimal lattice Green
function that is designed to minimize root-mean-square (rms) errors
\cite{deserno98a, sagui99}.  The PME and SPME algorithms have already
been generalized to compute dipolar interactions \cite{toukmaji00a}.
In this paper we perform a similar generalization, but for the \PPPM\
algorithm. An advantage of the \PPPM\ approach is that it provides
theoretical estimates for the rms accuracy of the forces, torques and
energy as by-products.  These estimates give valuable information
about the accuracy of the algorithm without having to perform tedious
benchmarking, and they allow for the tuning of the algorithm to yield
minimal computing time at a given level of accuracy.  No such
theoretical error estimates are currently known for the dipolar PME
nor SPME methods.
 
 To verify the applicability and correctness of the method presented in
 this article and to be able to perform the numerical tests, the method
 was implemented in the simulation package
 \es\cite{limbach06a,espressoweb}, and it will be contained in one of
 the coming releases of the software. 

The outline of this paper is as follows. The basic formulas for the
Ewald summation of dipolar interactions are recalled in
Sct.~\ref{sectionDipolarEwald}. In Sct.~\ref{sectionDipolarP3M},
Hockney and Eastwoods's \PPPM\ algorithm is extended to compute
dipole-dipole interactions. A correction term that must be applied to
any dipolar energy when computed via particle-mesh-methods is derived
in Sct.~\ref{Sct_Madelung}. Theoretical estimates for the rms error of
forces, torques, and energy as computed by \PPPM\ are presented in
Sct.  \ref{eefft}.  Numerical tests of the accuracy of the error
estimates are made in Sct.~\ref{sectionNumericalTest}.  
In Sct.~\ref{performance} several issues related to the computational 
efficiency of the method are discussed:  performance of the method when 
compared to the traditional dipolar Ewald sums,  suitable approaches to
make a fast implementation of the method in constant-pressure simulations, and 
a comparison of the efficiency of dipole-based and charge-based
models  to mimic  dipolar systems.
 Technical details for building up the \PPPM\
dipolar method are given in App.~\ref{superappendixA}, while
App.~\ref{superappendixB} derives and discusses the rms error
estimates.


\section{The dipolar \PPPM\ method}

In this section the dipolar \PPPM\ algorithm is presented by first
recalling the basics of the dipolar Ewald summation in which the new
method has its roots (see Sct. \ref{sectionDipolarEwald}).  The
details of the new algorithm are presented in Sct.
\ref{sectionDipolarP3M}. The effect of discretization errors in
Madelung-Self interactions (those of a particle with its periodic
images and itself) is discussed and a correction term to remove a bias
in the energy is derived in Sct. \ref{Sct_Madelung}. The different
Fourier Transforms as well as the domains to which they apply are
defined in Table I.
 In the following, we assume a cubic
box, but the generalization to triclinic boxes is straightforward, see
for instance ref. \cite{toukmaji00a} for an implementation in PME and
SPME algorithms.

\subsection{Ewald summation with dipolar interactions}
\label{sectionDipolarEwald}

The fundamental idea of the Ewald summation (and its advanced
implementations like the particle mesh methods PME, SPME and \PPPM) is
to calculate energies, forces, and torques by splitting the
long-ranged dipolar pair-interaction into two parts,
\begin{equation}
  v(\br,\bmu_i ,\bmu_j) = \left(\bmu_i \dotprod \nabla_{\br_i} \right) \left(\bmu_j \dotprod \nabla_{\br_j} \right) 
  \big(
  \psi(\br_{ij} )
  +\phi(\br_{ij}) \big),
\end{equation}
where $\psi(\br)$ contains the short-distance part of the Coulomb
interaction, and $\phi(\br)$ contains its long-distance part
($\phi(\br)$ must moreover be smooth everywhere and regular at the
origin). The standard way to perform this splitting is to set
\begin{gather}
\label{psi}
\psi(\br) \equiv \frac{\erfc(\alpha r)}{r}, \qquad r = |\br|,
\\
\label{phi}
\phi(\br) = \frac{\erf(\alpha r)}{r},
\end{gather}
though other choices are
possible~\cite{ewald21,rajagopal94,hunenberger00a,batcho01a}.  The
inverse length~$\alpha$, which is often referred to as the Ewald (or
splitting) parameter, weighs the importance of one term with respect
to the other, and can be chosen so as to optimize the performance. The
interactions associated to the function $\psi$ are short-ranged and
they can hence efficiently be summed numerically. The interactions
associated to the function $\phi$ are long-ranged in real space, but
short-ranged in the reciprocal Fourier space, and can therefore be
efficiently computed in that latter space.  The decomposition of the
potential leads to the well-known Ewald formula for the electrostatic
energy of a system of dipoles (see details in refs
\cite{weis05a,frenkel02a,deLeeuw80a,deLeeuw80b})
\begin{equation}
\label{U_Ewald}
U = U^{(r)}+ U^{(k)} + U^\self + U^\surf
\end{equation}
where the real-space energy $U^{(r)}$, the reciprocal-space energy
$U^{(k)}$, the self-energy $U^\self$ and the surface $U^\surf$
contributions are
\begin{gather}
\label{U^r}
U^{(r)} = \frac{1}{2} 
        \sum_{i,j=1}^{N} \sideset{}{^{'}}\sum_{\bn \in \mathbb{Z}^3}
(\bmu_i \dotprod \nabla_{\br_i}) \left(
\bmu_j \dotprod \nabla_{\br_j} \right) \psi(\br_{ij})
\\
\label{U^k}
U^{(k)} = \frac{1}{2V}\sum_{\bk\neq0 \atop \bk \in \mathbb{K}^3} |\FSp{\brho}(\bk)\dotprod i\bk|^2 \FTnp{\phi}(\bk)\\
U^\self = - \frac{2\alpha^3}{3\sqrt{\pi}} \sum_{i=1}^N \mu_i^2 \label{Selfenergia} \\
U^\surf = \frac{2 \pi}{(2\epsilon'+1)V} \sum_{i=1}^N
                   \sum_{j=1}^N\bmu_i  \dotprod\bmu_j,  \label{Surfenergia}
\end{gather}
where $V=L^3$ is the volume of the box, and $\epsilon'$ is the
dielectric constant of the medium surrounding the replica boxes:
$\epsilon'=1$ for vacuum, and $\epsilon'=\infty$ for metallic boundary
conditions.  Because of the periodic boundary conditions, wave vectors
$\bk \in {\mathbb{K}}^3$ are discrete where ${\mathbb{K}}^3 \equiv \{
2 \pi \bn/L : \bn \in {\mathbb{Z}}^3\}$.
In Eq.~\eqref{U^k}, $\FSp{\brho}(\bk)$
is the Fourier transform of the periodic dipole density 
\begin{equation}
\label{rhor}
\brho(\br) = \sum_{i=1}^N \bmu_i \,\delta(\br-\br_i), \qquad \br \in V,
\end{equation}
which reads, 
\begin{equation}
\label{rhok}
\FSp{\brho}(\bk) \equiv \mathrm{\FSpletters}[\brho](\bk) = \sum_{i=1}^N \bmu_i\, e^{-i\bk\cdot\br_i}.
\end{equation}
In~\eqref{U^k}, the Fourier transform $\FTnp{\phi}(\bk)$ of the
reciprocal interaction~\eqref{phi} is
\begin{equation}
\label{phik}
\FTnp{\phi}(\bk) =  \int\!\phi(\br) e^{-i\bk\cdot\br}
\dd\br = \frac{4\pi}{k^2}  e^{-k^2/4\alpha^2}.
\end{equation}

The term $U^\self$ subtracts the unwanted self-energies that are
included in the reciprocal energy $U^{(k)}$, where the self-energy of a
dipole is defined as the reciprocal interaction of the dipole with
itself: $\lim_{\bm{r } \rightarrow 0}(-\frac{1}{2}) \left( \bmu_i
  \cdot \nabla_{\br} \right)^2\phi(\br)$. 
It should be remarked that the expression given in eq.~\eqref{Surfenergia} for
the surface term is valid only when a spherical order of summation
is used in the calculation of the direct 
sum\cite{deLeeuw80a,deLeeuw80b}, eq.~\eqref{direct_sum}. In that case, 
eqs.~\eqref{direct_sum} and eq.~\eqref{U_Ewald} lead to identical values, 
provided that the interaction energy of the dipoles with the surrounding 
medium of dielectric constant $\epsilon'$ at infinity is added to 
eq.~\eqref{direct_sum} ($\epsilon'=1$ was assumed when writing 
\eqref{direct_sum}). Notice that the surface term  vanishes if 
metallic boundary conditions ($\epsilon'= \infty$) are used.

Ewald expressions for the force and electric field acting on a
dipole~$i$ follow from eqs. \eqref{eq3}, \eqref{eq4} and
\eqref{U_Ewald}:
\begin{gather}
\label{Ewald F_i}
\bm{F}_i = \bm{F}^{(r)}_i + \bm{F}^{(k)}_i \\
\label{Ewald E_i}
\bm{E}_i = \bm{E}^{(r)}_i + \bm{E}^{(k)}_i + \bm{E}^\self_i + \bm{E}^\surf_i.
\end{gather}
where the superscripts $(r)$ and $(k)$ denote the real-space and
reciprocal-space contributions.  Notice that there is no self- nor
surface-contribution to the force because the self- and surface-energy
terms (eqs.~\eqref{Selfenergia} and \eqref{Surfenergia}) are
independent of the particle positions.  By~\eqref{torques}, the torque
on dipole~$i$ follows directly from the electric field: $\btau_i =
\bmu_i\times \bm{E}_i$. The reader is referred to ref.~\cite{wang01a}
for fully explicit Ewald formulas for the real space and reciprocal
space contributions to the force and torque. For further reference, it
is worth noting that the reciprocal space contributions to the force
and electrostatic field can be written as
\begin{eqnarray}
\label{E^k}
\bm{E}^{(k)}_i &=&   {\FSpletters}^{-1}_{\bk \ne 0} \left[   \FSp{\bm{E}}^{(k)}  \right]  = {\FSpletters}^{-1}_{\bk \ne 0} \left[ i\bk
 \left(\FSp{\brho}(\bk)\dotprod i\bk \right)
\FTnp{\phi}(\bk) \right]  \\ \notag
\bm{F}^{(k)}_i & =& {\FSpletters}^{-1}_{\bk \ne 0} \left[
i\bk (\bmu_i \cdot i\bk)  \left(\FSp{\brho}(\bk)\dotprod i\bk \right)
\FTnp{\phi}(\bk) 
\right] \\ 
\label{F^k}
\phantom{\bm{F}^{(k)}_i} &=& 
 \, \mu_{i,x} \, {\FSpletters}^{-1}_{\bk \ne 0} \left[ \FSp{{E}}_x^{(k)}~ i \bk \right] 
 \,+ \mu_{i,y} \, {\FSpletters}^{-1}_{\bk \ne 0} \left[ \FSp{{E}}_y^{(k)} ~ i \bk \right] 
 \,+ \mu_{i,z} \, {\FSpletters}^{-1}_{\bk \ne 0} \left[ 
 \FSp{{E}}_z^{(k)} ~ i \bk \right] \\   \notag
 &=&
  \, \mu_{i,x} \, {\FSpletters}^{-1}_{\bk \ne 0} \left[ \FSp{\bm{E}}^{(k)}~  i  k_x \right] 
 \,+ \mu_{i,y} \, {\FSpletters}^{-1}_{\bk \ne 0} \left[ \FSp{\bm{E}}^{(k)} ~ i k_y \right] 
 \,+ \mu_{i,z} \, {\FSpletters}^{-1}_{\bk \ne 0} \left[  \FSp{\bm{E}}^{(k)} ~ i k_z \right] 
\end{eqnarray}
where the inverse Fourier series $\mathrm{\FSpletters}^{-1}[\cdots]$ is defined in Table~I (the $\bk=0$ term must be excluded in the
back-transformation), and the components of the Fourier transform of the
electrostatic field are  $  \FSp{\bm{E}}^{(k)} = (\FSp{{E}}_x^{(k)},\FSp{{E}}_y^{(k)} , \FSp{{E}}_z^{(k)}) $, and 
 $\bmu_i  = (\mu_{i,x}   \,  ,  \mu_{i,y}  \,   ,  \mu_{i,z})$ is the
 dipole moment of particle~$i$. The last equality for the force arises
 from the fact that  $\bm{F}(\br) = \nabla_{\br} (\bmu \dotprod \bm{E}(\br))  =  (\bmu \dotprod  \nabla_{\br} ) \bm{E}(\br)$ in electrostatics ($\nabla \times \bm{E}=0$).

 From a computational point of view, the Ewald
method requires therefore to first Fourier transform the dipole
distribution to the reciprocal space, then to solve the Poisson
equation in reciprocal space [which reduces to a simple multiplication
by~$\FTnp{\phi}(\bk)$], and finally to Fourier-back-transform the
results to real space.

 
\subsection{Algorithmic details of the mesh calculations}
\label{sectionDipolarP3M}

What distinguishes the particle mesh methods from the Ewald summation
is that, while Ewald summation uses the standard Fourier series to
compute the reciprocal space contribution, particle mesh methods use
Fast Fourier Transformations ($\FFT$), thereby reducing the
computational effort from ${\cal O}(N^{3/2})$ to ${\cal O}(N \log N)$.
However, since $\FFT$ is a mesh transformation, it is necessary to:
(1) Map the dipole moments from continuous positions onto lattice
points (which will be referred to as \emph{dipole assignment} to the
mesh sites); (2) Fast-Fourier transform the mesh and solve the Poisson
equation on the (reciprocal) mesh; (3) Fourier transform the mesh back
to real-space, and interpolate the results onto the continuous dipole
positions.

The computation of the real-space contribution $U^{(r)}$ in the Ewald
formula is kept unchanged, and the reader is referred to
\cite{wang01a} for explicit formulas. In the following, we discuss in
detail the mesh calculation in the case where the
$i\bk$-differentiation scheme is used.  Other differentiation schemes
can be easily implemented, see \cite{deserno98a} for details.

The mesh is assumed to be a cubic FFT mesh with the lattice spacing
given by $h=L/N_M$, where $N_M$ stands for the number of mesh points
in each direction. We denote by ${\mathbb{M}}^3$ the set of all points
belonging to the mesh: ${\mathbb{M}}^3 \equiv \{ \bn h : \bn \in
{\mathbb{Z}}^3, 0 \leq n_{x,y,z} < N_M \}$. An index `M' is attached
to any quantity defined at mesh points only, \eg\ the mesh-based
dipole density $\rho_{M}(\br_m)$ or the mesh-based electric field
$\bm{E}_M(\br_m)$, $\br_m\in\mathbb{M}^3$.  The inverse fast Fourier
transform FFT$^{-1}[\FFTs{f}]$ corresponds to a truncated Fourier
series over wave vectors in one Brillouin zone (see
Table I
). By convention, this zone is taken to be the set
of wave vectors $ \FFTs{\mathbb{M}}^3 \equiv \{ 2\pi {\bm n}/L : \bn
\in {\mathbb{Z}}^3, |n_{x,y,z}| < N_M/2 \}$, which we call the
``reciprocal mesh'' or first Brillouin zone. The number of mesh points
per direction $N_M$ should preferably be a power of two, because in
that case the FFTs are computed more efficiently. Notice that with
this definition, the reciprocal mesh is symmetric: if wave vector
$\bk$ belongs to the mesh, so does $-\bk$.

\subsubsection{Dipole assignment}
\label{dipole-assignement}

The dipole density $\brho_M(\br_m)$ on the mesh is determined from the
$N$ dipolar particles $\{(\br_i,\bmu_i)\}$ by the \emph{assignment
  function} $W(\br)$ that maps the particles from their continuous
positions to the mesh,
\begin{equation}
\ \brho_M(\br_m) = \frac{1}{h^3} \sum_{i=1 \atop \mic}^{N} \bmu_{i}
W(\br_m-\br_i),
 \label{mesh_density}
\end{equation}
where minimum image convention (\mic)  is used when computing relative distances $\br_m - \br_i$.
We use the same
assignment functions $W(\br)$ as defined by Hockney and Eastwood in
the original \PPPM\ method for Coulomb interactions\cite{hockney81a},
which are (shifted) B-splines and are tabulated in
ref.~\cite{deserno98a}. The assignment functions are classified
according to the number~$P$ of nearest grid points per coordinate
direction over which the dipole is distributed.  The quantity $P$ is
referred to as the \emph{assignment order parameter}. A formal
expression for Hockney and Eastwood's assignment functions is
$W^{(P)}({\bm r})=W^{(P)}(x)W^{(P)}(y)W^{(P)}(z)$ where
\begin{equation}
  W^{(P)}(x) = \underbrace{
    \left( \chi[\frac{-1}{2},\frac{1}{2}] \star ... \star   \chi[\frac{-1}{2},\frac{1}{2}]
    \right)
  }_{P\mathrm{-fold-convolution}} \left(
    \frac{x}{h} \right)
 \label{assignment1}
\end{equation}
and $\chi[\frac{-1}{2},\frac{1}{2}]$ is the characteristic function,
\ie{}, the function that is $1$ within this interval and $0$ outside.

\subsubsection{Solving the Poisson equation}
\label{solvPoissonequ}

The reciprocal electrostatic energy, and electrostatic field  are
computed at each mesh point $\br_m$ by approximating equations
\eqref{U^k},  and \eqref{E^k}  by
\begin{gather}
  \label{KenergiaBiased} 
  U^{(k)}_{M}  =  \frac{1}{2 V} \sum_{ \bk \in  \FFTs{\mathbb{M}}^3  \atop \bk\neq 0}  
  \left| \FFTs{\brho}_M(\bk)   \dotprod \FFTs{\bm{D}}(\bk)  \right|^2 
  \FFTs{G}(\bk) \\
  \label{mesh-electrostatic-field}
  \bm{E}^{(k)}_{M}(\br_m)  = 
  \FFT^{-1}_{\bk \ne 0} \left[ \FFTs{ \bm{E}}^{(k)}_{M}  \right]    
   =\FFT^{-1}_{\bk \ne 0} \left[
    \FFTs{\bm{D}}(\bk) \left(
     \FFTs{\brho}_M(\bk) \dotprod \FFTs{\bm{D}}({\bm  k})
     \right)   \FFTs{G}(\bk) \right] 
     (\br_m) .  
 \end{gather}
Here, $\FFTs{\brho}_M(\bk)$ is the fast Fourier transform of the
dipole density $\brho_M(\br)$ on the mesh.  The $\bk=0$ term is
excluded in the inverse transform FFT$^{-1}$ of  all  mesh-based
quantities as in reciprocal Ewald terms eqs.~\eqref{U^k},
\eqref{E^k}, and \eqref{F^k}.  
 The function
\begin{equation}
\label{definingD}
\FFTs{\bm{D}}(\bk)=i\bk, \qquad \bk \in \FFTs{\mathbb{M}}^3,
\end{equation}
is the Fourier expression of the gradient operator on the reciprocal
mesh. $\FFTs{G}(\bk)$ is the lattice Green function, also known as
the \emph{influence function}, and it is defined below at the end of Sec.~B [see eq.~\eqref{optimized_G}]. It should be remarked that both 
$\FFTs{\bm{D}}(\bk)$ and $\FFTs{G}(\bk)$
 are periodic in $\FFTs{\mathbb{K}}^3$, with the period given by the 
 first Brillouin cell 
 $\FFTs{\mathbb{M}}^3$,  \ie{}, period $2\pi/h$.

Note that eqs.~\eqref{KenergiaBiased} to
\eqref{mesh-electrostatic-field} correspond to the reciprocal Ewald
formulas recalled in Sct.~\ref{sectionDipolarEwald}, but are modified
in two ways: the $\FSpletters$ of the dipole density is replaced by a
$\FFT$ of the mesh dipole density and the (continuous) reciprocal
interaction $\FTnp{\phi}(\bk)$ is replaced by a discrete lattice Green
function $\FFTs{G}(\bk)$. A fundamental idea in the \PPPM\ method is
that the lattice Green function is not simply taken as the continuum
Green function $\FTnp{\phi}(\bk)$, but it is considered as an
adjustable function whose form is determined by the condition that the
mesh based calculation gives results as close as possible, in a
least-square sense, to the results of the original continuum problem (see
below Sec. \ref{GreenfunctionSec} for more details).

\subsubsection{Back-interpolation}

The mesh based electrostatic field is finally interpolated
back to the particle positions $\bm{r}_i$ (and possibly also to any
other point in the simulation box) using the same assignment function
$W(\br)$ and the minimum image convention (\mic):
\begin{eqnarray}
\bm{E}^{(k)}(\br_i) & = & \sum_{\bm{r}_m \in \mathbb{M}^3 \atop \mic}
\bm{E}^{(k)}_M(\br_m)W(\br_m-\br_i), \label{mapbackE} 
\end{eqnarray}
Once the electric field is known, the torques are obtained by
eq.~\eqref{torques} and the electrostatic energy of dipole~$i$, is given by
\begin{eqnarray}
U_i^{(k)} =-~ \bmu_i \dotprod  \bm{E}^{(k)}(\br_i).
\end{eqnarray}
Note that if only the total electrostatic energy is needed, it can be
obtained via eq.~\eqref{KenergiaBiased} which does not need any
inverse Fourier transform nor back-interpolation.

The force acting onto a particle $i$ can be obtained by analogy with
eq.~\eqref{F^k} as
\begin{eqnarray}
\bm{F}^{(k)}(\br_i) & = & \sum_{\bm{r}_m \in \mathbb{M}^3 \atop \mic} W(\br_m-\br_i)
\left\{
\; \mu_{i,x} \;   \FFT^{-1}_{\bk \ne 0} \left[  \FFTs{E}^{(k)}_{M,x} \FFTs{\bm{D}}({\bm  k})\right] +
\right.  \nonumber \\ 
& & \left.
\; \mu_{i,y} \;   \FFT^{-1}_{\bk \ne 0} \left[ \FFTs{E}^{(k)}_{M,y} \FFTs{\bm{D}}({\bm  k}) \right]+
\; \mu_{i,z} \;   \FFT^{-1}_{\bk \ne 0} \left[ \FFTs{E}^{(k)}_{M,z} \FFTs{\bm{D}}({\bm  k}) \right]
\right\},
\label{mapbackF}
\end{eqnarray}
where  the reciprocal mesh electrostatic field is
$\FFTs{\bm{E}}^{(k)}_{M}=( \FFTs{E}^{(k)}_{M,x}  \,  , 
\FFTs{E}^{(k)}_{M,y}   \,  , \FFTs{E}^{(k)}_{M,z}  )$. In the last
formula the differential operator and the electrostatic field can be
permuted as in eq.~\eqref{F^k} .

The differentiation used in step 2 and in eq.~\eqref{mapbackF} (the so-called $i\bk$-\emph{differentiation} or force-interpolation scheme which consists in multiplying the reciprocal mesh by $\FFTs{\bm{D}}(\bk)=i\bk$) is the
most accurate variant when combined with the assignment scheme
employed in section \ref{dipole-assignement}. Note, however, that to
compute the forces and electric field vectors, it requires the
back-FFT of vectorial quantities.  By contrast, in the analytical differentiation
scheme as used in the SPME algorithm, the forces and electrical field
vectors are derived in real space from the back-transformed potential mesh with the
subsequent saving of FFT's. 
Analytical differentiation leads however to forces that violate Newton's third law and hence that do not conserve momentum. A global correction can be applied to restore conservation  of the total momentum, but its effects on the physics of the system is difficult to assess. An algorithm that uses analytical differentiation without introducing such spurious forces is currently under study.

\subsubsection{The lattice Green function}
\label{GreenfunctionSec}

The optimal lattice Green function to compute dipolar interactions can
be found by minimizing the rms error in the (reciprocal) pair
interaction $\bT^{(k)}$ between two unit dipoles in the simulation
box:
\begin{multline}
  \label{HE-error-measure}
  Q^2_\INT[\bT^{(k)}]:=\frac{1}{h^3 (4 \pi)^2 V}
  \int_{h^3}\dd\br_1 \int_{V}\dd\br_2
  \int\dd\bm{\Omega}_1
  \int\dd\bm{\Omega}_2  \\
  \left[ \bT^{(k)}(\br_1,\hat{\bmu}_1,\br_2,\hat{\bmu}_2 ) 
    - \bT^{(\mathrm{ex}, k)}(\br_1,\hat{\bmu}_1,\br_2,\hat{\bmu}_2) \right]^2
\end{multline}
where $\bT^{(\mathrm{ex}, k)}(\br_1,\bmu_1,\br_2,\bmu_2)$ is the exact
(reciprocal) dipolar Ewald interaction (energy, electrostatic field,
force or torque) between two dipoles, and
$\bT^{(k)}(\br_1,\bmu_1,\br_2,\bmu_2)$ is the \PPPM\ pair
interaction. The quantity $Q_\INT^2$ defined in~\eqref{HE-error-measure} is
the squared error of the \PPPM\ interaction averaged over all
positions and orientations of the two dipoles in the simulation box.
Notice that the average over $\br_1$ has been restricted to a single
mesh cell $h^3$ thanks to the periodicity of the system.

The optimal influence function which result from the minimization of
eq. \eqref{HE-error-measure} is found to be (see
App.~\ref{superappendixA})
\begin{equation}
  \FFTs{G}_\opt(\bk) = \frac{ \displaystyle \sum_{\bmm \in \mathbb{Z}^3} \left[
      \left[ \FFTs{\bm{D}}(\bk) \dotprod i {\bmeta}_{\bmm} \right]^S 
      \left(\FTnp{U}(\bmeta_{\bmm}) \right)^2 
      \FTnp{\phi}(\bmeta_{\bmm}) \right]} 
  {\displaystyle  \left[ \FFTs{\bm{D}}(\bk)   \right]^{2S} 
    \left[ \sum_{\bmm \in \mathbb{Z}^3}
      \left(\FTnp{U}(\bmeta_{\bmm})\right)^2 \right]^2}
  \label{optimized_G} 
\end{equation}
where $\bmeta_{\bmm} \equiv \bk + \left( 2\pi/h \right) \bmm$, $
\FTnp{U}(\bk) \equiv \FTnp{W}(\bk)/h^3$, and $\FTnp{W}(\bk)$ is the
Fourier transform of the assignment function defined in
eq. \eqref{assignment1},
\begin{equation}
  \FTnp{W}(\bk) = h^3 \left(    
  \frac{\sin(\frac{1}{2}k_xh)\sin(\frac{1}{2}k_yh)\sin(\frac{1}{2}k_zh)}
  {(\frac{1}{2}k_xh)(\frac{1}{2}k_yh)(\frac{1}{2}k_zh)}  
   \right)^P.
\end{equation}

The influence function for dipolar forces is obtained by setting $S=3$
in the previous expression.  The value $S=2$ refers to the optimal
influence function for the dipolar torques, energy, and the
electrostatic field.

The form of these influence functions resembles the influence function
obtained by Hockney and Eastwood for Coulomb forces ($S=1$). It should
be remarked that the use of the different influence functions to
compute the forces and torques does not imply any noticeable time
overhead because influence functions are computed and stored at the
beginning of the simulation, and they remain unaltered throughout the
whole simulation.

When implementing the method, it is important that the reciprocal mesh
is symmetric to avoid systematic biases on the computed quantities
(see App.~\ref{superappendixB}) \cite{commentsymmetrization}.


\subsection{Madelung-Self interactions and correction term for the energy}
\label{Sct_Madelung}

Fast-Fourier-transforms greatly accelerate the calculation of the
Ewald reciprocal interactions, but have the drawback of introducing
discretization errors in the computed quantities. On the one hand,
these errors arise from truncation of the Fourier series, as wave
vectors greater than $2\pi/h$ are discarded in the mesh calculation,
and on the other hand from aliasing, which is caused by band-folding
in Fourier space due to undersampling of the continuous dipole
distribution~\cite{hockney81a}. The discretization errors do not
necessarily average to zero, so \PPPM\ quantities may be biased. This
is the case for the reciprocal energies computed on the mesh, which
need hence to be corrected by applying a shift which is determined
below [eq.~\eqref{Ukws}]. No similar correction needs to be applied to
\PPPM\ forces and torques.

\subsubsection{Madelung-Self interaction}

The bias in the \PPPM\ energies originates from the fact that the
Madelung and self interactions are not fully accounted for in the mesh
calculation.   For Coulomb interactions, the issue has been discussed in
detail by H\"unenberger \cite{hunenberger02a} and Ballenegger et al.
\cite{ballenegger08a}.
The exact Madelung interaction (energy, force or torque)
is defined as the interaction of a dipole with all its images in the
periodic replicas of the simulation box:
\begin{equation}
\label{U_Madelung(mu)}
   U_{\mathrm{Madelung}}^\exact({\bmu}) = \frac{1}{2} \sum_{\bm{n}\in \mathbb{Z}^3 \atop \bm{n} \neq 0} v(\bm{n} L, {\bmu},{\bmu})
\end{equation}
where the sum over images must be performed in concentric shells and
the vacuum boundary condition ($\epsilon'=1$) is employed
in~\eqref{U_Madelung(mu)}. The Madelung energy depends only on the
dipole moment ${\bmu}$ and the length $L$ of the cubic simulation
box. Due to the specific form of the dipolar
interaction~\eqref{v_dip}, the sum in~\eqref{U_Madelung(mu)} vanishes,
as proved by de Leeuw et al. \cite{deLeeuw80a}. Consequently, the
exact Madelung dipolar energy, force and torque are zero. Notice that
the use of the Ewald summation~\eqref{U_Ewald} to compute the Madelung
energy~\eqref{U_Madelung(mu)} leads to the relation
\begin{equation}
\label{JuanTrick}
   U_{\mathrm{Madelung}}^{(\mathrm{ex},r)}({\bmu})
+ U_{\mathrm{Madelung}}^{(\mathrm{ex},k)}({\bmu}) - \frac{2 \alpha^3\mu^2}{3\sqrt{\pi}}
+ \frac{2\pi\mu^2}{3 L^3} = 0.
\end{equation}
However, if this energy is computed with the \PPPM\ algorithm, for
example by putting a single dipolar particle in the simulation box,
the obtained energy $U(\br,\bmu)$ differs from zero because the
dipolar interactions with the images of the dipole are only
approximately accounted for.  Furthermore, the (reciprocal)
interaction of the dipole with itself, which is included in the mesh
calculation of $U^{(k)}(\br,\bmu)$, is also only approximately
accounted for because of the discretization errors.  The later
subtraction of the exact self-energy by the term $U^\self = - {2
  \alpha^3}\mu^2/({3\sqrt{\pi}})$ will therefore not exactly
compensate the unwanted self-interaction. These two effects are
responsible for a systematic bias in the \PPPM\ energies because the
discrepancy between the exact and \PPPM\ values does not vanish on
average.  We call the sum of the Madelung and self-interaction the
``Madelung-Self'' (MS) interaction. More precisely, it is defined as
the sum of the direct and reciprocal space contribution to the energy
(or force or torque) in a one particle system, namely
\begin{equation}
\label{VB_U_ms}
U_\MS(\br,\bmu) \equiv  U^{(r)}_{\mathrm{Madelung}}(\br,\bmu) + U^{(k)}_{\mathrm{Madelung}}(\br,\bmu)
\end{equation}
(with this definition, $U_\MS$ is independent of the choice of the
boundary condition~$\epsilon'$). Contrary to the exact \madelungself\
energy, which reads, from~\eqref{JuanTrick},
\begin{align}
\label{U_MS^ex}
U_\MS^{\exact}({\bmu}) &= U_{\mathrm{Madelung}}^{(\mathrm{ex},r)}({\bmu})
+ U_{\mathrm{Madelung}}^{(\mathrm{ex},k)}({{\bmu}}) \notag \\ &= \mu^2
\left( \frac{2 \alpha^3}{3\sqrt{\pi}} - \frac{2\pi}{3 L^3} \right),
\end{align}
the \madelungself\ energy in \PPPM\ \eqref{VB_U_ms} depends in general
both on the position and on the orientation of the dipole moment
because of the mesh calculation.

\subsubsection{Correction term for the \PPPM\ energy}
The error in the \PPPM\ energy of a dipolar particle located at $\br$
with dipole moment $\mu$ in direction $\hat{\bmu}$ is
\begin{equation}
\label{VB_DeltaU}
   \Delta U(\br,\bmu) = \mu^2\, ( U_\MS(\br,\hat{\bmu}) - U_\MS^{(ex)}(\hat{\bmu}) ),
\end{equation} 
where we factored out the magnitude $\mu^2$.  This error does not
vanish when averaged over all positions and orientations of the
dipolar particle. The sum of these average errors for all dipoles
$\{\bmu_i\}_{i=1,...,N}$ provides the correction term
\begin{equation}
\label{Ukws}
  \bracket{U^\corr} = -M^2 \bracket{\Delta U(\br,\hat{\bmu})}
\end{equation}
that must be added to the \PPPM\ energies to remove the bias (at least
on average).  In eq.~\eqref{Ukws},
\begin{equation}
   M^2 \equiv  \sum_{i=1}^N \mu_i^2,
\end{equation}
and the average error $\bracket{\Delta U(\br,\hat{\bmu})}$ is easily
determined analytically. Indeed, we have
\begin{equation}
\label{average_error}
   \bracket{\Delta U(\br,\hat{\bmu})} =  \bracket{U_\MS^{(k)}(\br,\hat{\bmu})} - 
\frac{2 \alpha^3}{3\sqrt{\pi}}
+ \frac{2\pi}{3 L^3} ,
\end{equation}
where we used \eqref{VB_DeltaU}, \eqref{U_MS^ex} and the fact that
there is no real-space contribution to the MS energy in the \PPPM\
calculation when the minimum image convention (m.i.c.) is used. The
average reciprocal-space MS energy is calculated in
App.~\ref{appendixD} and reads
\begin{equation}
  \bracket{U_{\MS}^{(k)}(\br,\hat{\bmu})} = 
  \frac{1}{6V} \sum_{ \bk \in \FFTs{\mathbb{M}}^3 \atop \bk \ne 0} 
   \FFTs{\bm{D}}^2(\bk)~
  \FFTs{G}(\bk) \sum_{\bmm \in
    \mathbb{Z}^3} \FTnp{U}^2(\bk_{\bmm} )
\label{Umean1particle}
\end{equation}
with $\bmeta_{\bmm} \equiv \bk +  \left(2\pi/h \right) \bmm$. 

In conclusion, the corrected formula for the \PPPM\ energy is
\begin{equation}
\label{P3M_energy}
U_\pppm = U^{(r)} + U^{(k)}_{M} + U^\self + U^\surf +\bracket{U^\corr}.
\end{equation}
Note that the correction term only needs to be computed once at the
beginning of the simulation, hence it is inexpensive in CPU cost, but
its usage can improve the accuracy of the dipolar \PPPM\ energies by
several orders of magnitude (\eg\ inset of Figure~\ref{f3}) depending
on the values of the mesh size $N_M$ and the Ewald splitting parameter
$\alpha$.

\subsubsection{Madelung-Self forces and torques}

Since each dipole in \PPPM\ is subject to a position- and orientation-
dependent MS energy $U_\MS(\br,\bmu)$, it can be expected from
relations \eqref{eq3}-\eqref{eq4} that it will also experience an MS
force and an MS torque. The \PPPM\ force is obtained from the mesh
using eq.~\eqref{mapbackF} (instead of eq.
\eqref{eq3}), and it is proved in App.~\ref{appendixD} that the MS
force cancels out.  Consequently, \PPPM\ conserves the momentum in
difference to SPME, for example. In the same appendix, it is also shown that a
non-vanishing MS torque does arise in the mesh calculation. However,
on average this MS torque vanishes and does therefore not result in a
systematic bias to the torques.

The results on \madelungself\ interactions are summarized in
Table II.
 The fluctuating errors in MS interactions have an
impact on the accuracy of the computed quantities. The rms error
estimates for \PPPM\ energies and torques are therefore more difficult
to obtain than the one for forces (see next section).

We stress that MS interactions are common to all particle mesh
methods, and the explicit expression for the possible biases (such as
the energy correction \eqref{Ukws}) depends on the details of each
algorithm. This is the first work, together with \cite{ballenegger08a}, in which the effect of the MS
interactions is thoroughly assessed in a particle-mesh method.

\section{Error estimates for the dipolar \PPPM~algorithm}
\label{eefft}
  
In this section, theoretical error estimates for the root-mean-square
(rms) error of the energy, forces and torques for the \PPPM\ algorithm
are presented.  The accuracy of the \PPPM\ method depends on the
chosen values for the parameters of the method: the Ewald splitting
parameter $\alpha$, the real-space cut-off distance $\rcut$, the mesh
size $N_M$ and the assignment order $P$, as well as on parameters of
the system: the number of particles $N$, the box length $L$ and the
sum over all  squared dipole moments, $M^2$.

It is very useful to have formulas that are able to predict the error
associated to a set of parameter values.  Not only do such formulas
enable the user to control the accuracy of the calculation, but they
also allow for an automatic tuning of the algorithm, so that it can
run at its optimal operation point, thus saving  computer time.

A measure of the accuracy is given by the rms error defined by
\begin{equation}
\Delta T \equiv  \bracket{\sqrt{ \frac{1}{N} \sum_{i=1}^{N} 
  \left( \bT(i)  - \bT^\exact(i) \right)^2 } } 
\label{def-deltaT}
\end{equation}
where $\bT(i)$ is the value of~$\bT$ (for example electrostatic field,
force, torque or energy) associated to particle~$i$ as obtained from
the \PPPM\ method, and $ \bT^\exact(i)$ is the exact value as defined
by the direct summation formulas (eqs. \eqref{direct_sum},
\eqref{eq3}, \eqref{eq4}). The angular brackets denote an average over
particle configurations.  In~\eqref{def-deltaT}, $i$ is a short-hand
notation for $(\br_i,\bmu_i)$. In the case where the \emph{total}
electrostatic energy $U$ is measured, the rms error is defined by
\begin{equation}
\Delta U \equiv \sqrt{\bracket{ \left( U - U^\exact \right)^2   } },
\label{uest1}
\end{equation}
where $U$ is the corrected \PPPM\ energy \eqref{P3M_energy}, and
$U^\exact$ is the exact energy \eqref{direct_sum}.

Eqs. \eqref{def-deltaT} and \eqref{uest1} are calculated analytically
in the App.~\ref{superappendixB} to get useful error estimates as
functions of the various parameters. The calculation is done under the
assumption that the positions and orientations of the dipoles are
distributed randomly.  In Sct.~\ref{sectionNumericalTest} it is shown
that our rms error estimates still accurately predict the errors for
dipolar systems in which the dipoles are strongly correlated. For
random systems, the average over configurations reduces to
\begin{equation}
\label{<>}
\bracket{ \cdots  }  \equiv \frac{1}{V^N}\frac{1}{(4\pi)^N} \int
\cdots \int \cdots \, \dd1\ldots\dd N
\end{equation}
where $\int\ldots\dd i$ denotes integration over all positions and
orientations of particle~$i$.

As shown in App.~\ref{superappendixB}, the rms error
arises from two distinct
contributions: errors in the interaction of a particle $i$ with a 
particle~$j\neq i$ (including the images of particles~$j$ in the
periodic replicas of the simulation box), and errors in the Madelung-Self 
interactions of each particle. The first contribution is denoted by the
subscript $\INT$, while the latter contribution is denoted by the
subscript $\MS$.  In App.~\ref{superappendixB}, the following three 
rms error estimates for the dipolar \PPPM\ method are
derived.

\subsection{Error in the dipolar forces}
The rms error estimate for dipolar forces is given by
\begin{equation}
  \left( \Delta F \right)^2 \simeq \big(\Delta F^{(r)}\big)^2 + \frac{M^4}{N} Q^2_\INT[F^{(k)}],
\label{summarydF}
\end{equation}
where $\Delta F^{(r)}$ is the real space error, \cite{wang01a}
\begin{eqnarray}
  \Delta F^{(r)}  &\simeq&  M^2  \left( V\alpha^4 \rcut^9 N \right)^{-1/2}
  [ \frac{13}{6}C_c^2+\frac{2}{15}D_c^2 - \frac{13}{15}C_cD_c ]^{1/2} e^{-\alpha^2
    \rcut^2}  \label{DeltaF^r}\\
\label{cc}
C_c&\equiv&4 \alpha^4\rcut^4 + 6 \alpha^2 \rcut^2 +3  \\
D_c&\equiv&8 \alpha^6\rcut^6 + 20 \alpha^4 \rcut^4 +30 \alpha^2 \rcut^2 + 15
\end{eqnarray}
and $Q^2_\INT[F^{(k)}]$ is given by the general expression
$Q^2_\INT[T^{(k)}]$ in which  the optimal influence function
$\FFTs{G}_\opt(\bk)$ is used, namely
\begin{equation} 
\label{Qopt}
Q^2_\INT[T^{(k)}] = \frac{a }{9V^2} \sum_{ \bk \in \FFTs{\mathbb{M}}^3
  \atop \bk \ne 0} \Bigg\{ \sum_{\bmm \in \mathbb{Z}^3} |\bk_{\bmm}
|^{2S} \big( \FTnp{\phi}(\bk_{\bmm} ) \big)^2  - \frac{
  \big(\sum_{\bmm \in \mathbb{Z}^3} \left( \FFTs{\bm{D}}(\bk) \dotprod
    i \bk_{\bmm} \right)^S \big(\FTnp{U}(\bk_{\bmm}) \big)^2
  \FTnp{\phi}(\bk_{\bmm}) \big)^2 } { \big( \FFTs{\bm{D}}(\bk)
  \big)^{2S} \Big[ \sum_{\bmm \in \mathbb{Z}^3}
  \big(\FTnp{U}(\bk_{\bmm})\big)^2 \Big]^2} \Bigg\},
\end{equation} 
using the parameters $(S=3,a=1)$ for dipolar forces. The short hand
notation $\bk_{\bmm} \equiv {\bm k}+\frac{2\pi}{h}\bmm$ is used.

\subsection{Error in the torques}

The rms error estimate for dipolar torques is
\begin{equation}
\left( \Delta \btau \right)^2 \simeq \big(\Delta \btau^{(r)}\big)^2 +
\frac{ M^4}{N} Q^2_\INT[\btau^{(k)}] +
 \frac{\sum_i \mu_i^4}{N} Q^2_\MS[\btau^{(k)}]
\label{summarydTau}
\end{equation}
where the real-space contribution $\Delta {\tau}^{(r)}$ is
\begin{equation}
\Delta {\tau}^{(r)} \simeq M^2 \left( V\alpha^4 \rcut^7 N \right)^{-1/2} 
[ \frac{1}{2}B_c^2+\frac{1}{5}C_c^2 ]^{1/2} 
e^{-\alpha^2 \rcut^2},
\label{DeltaT^r}
\end{equation}
with $B_c \equiv 2 \alpha^2 \rcut^2 +1$
%
and $Q^2_\INT[\btau^{(k)}]$ is given by \eqref{Qopt} using
$(S=2,a=2)$.   The expression for  $Q^2_\MS[\btau^{(k)}]$ reads
\begin{eqnarray}
 \label{dtw2} 
Q^2_\MS[\btau^{(k)}]  &=& 
\frac{1}{6 V^2}
\sum_{ \bkone \in \FFTs{\mathbb{M}}^3 \atop \bkone \ne 0} \sum_{\bkprima
  \in \FFTs{\mathbb{M}}^3 \atop \bkprima \ne 0} 
\FFTs{G}(\bkone ) ~ \FFTs{G}(\bkprima )~  
h( \FFTs{\bm{D}}(\bkone), \FFTs{\bm{D}}(\bkprima) )~
\\ \nonumber
& &
\sum_{\bm{t} \in \mathbb{Z}^3} \sum_{\bm{l} \in \mathbb{Z}^3} \sum_{\bmm \in \mathbb{Z}^3}    \left[
  \FTnp{U}(\bkone_{\bm{t}} ) \FTnp{U}(\bkprima_{\bm{l}}) ~
  \FTnp{U}( \bkone_{\bm{tm}} ) ~
  \FTnp{U}( \bkprima_{\bm{lm}} )  \right]
 \end{eqnarray}
where 
\begin{eqnarray}
\label{hequation}
h( \bm{a}, \bm{b} ) & \equiv & \left[ 
2 \left( \bm{a} \dotprod  \bm{b} \right)^2 - \frac{1}{5} \left(
\frac{| \bm{a} + \bm{b } |^4+| \bm{a} - \bm{b }  |^4}{2}-\bm{a}^4 -\bm{b}^4 \right)
 \right] 
\end{eqnarray}
and $\bmeta_{\bm{\alpha}} \equiv \bk + (2\pi/h)
\bm{\alpha}$ , $\bmeta_{\bm{ \alpha \beta}} \equiv
\bk +(2\pi/h) (\bm{\alpha}+\bm{\beta})$.

The expression in eq. \eqref{dtw2} is certainly cumbersome, it
involves a 15-fold sum which renders the expression difficult to
evaluate.  A very easy way to substantially reduce the time needed to
compute eq. \eqref{dtw2} is to skip the inner loops whenever their
maximal value is smaller than a desired accuracy.  An additional
reduction in the computer time by roughly a factor $64$ can be
obtained if one takes into account that aside of the function $h(
\FFTs{{\bm D}}(\bkone), \FFTs{\bm{D}}(\bkprima) )$, the remaining
coefficients are symmetric with respect to the sign inversion of each
one of the components of the vectors $\bkone$ and $\bkprima$.  In
fact, it is shown in Sct. \ref{sectionNumericalTest}, that in practice
the optimal performance point can be located with sufficient accuracy
when $\frac{\sum_i \mu_i^4}{N} Q^2_\MS[\btau^{(k)}]$ is completely
neglected in eq. \eqref{summarydTau}.

\subsection{Error in the total energy}
\label{totalenergyerror}
The rms error estimate for the total dipolar energy is 
\begin{equation}
  \left( \Delta U \right)^2 \simeq \big(\Delta U^{(r)}\big)^2 +
  2  M^4 Q^2_\INT[U_{nc}^{(k)}]
  +    \bracket{ \left( \Delta U_{nc,\MS}^{(k)}\right)^2}                               
  -   \left(\bracket{U^\corr}  \right)^2,
\label{summarydU} 
\end{equation}
where $U_{nc}$ is the non corrected energy [obtained by dropping
$\bracket{U^\corr}$ in \ref{P3M_energy}]. The real-space contribution
$\Delta U^{(r)}$ is
\begin{equation}
\Delta U^{(r)} \simeq M^2 \left(V \alpha^4 r_{\mathrm{cut}}^7 \right)^{-1/2} 
[\frac{1}{4}B_c^2+\frac{1}{15}C_c^2 -\frac{1}{6}B_c C_c ]^{1/2}  e^{
-\alpha^2 r_{\mathrm{cut}}^2 }.
\label{DeltaU^r}
\end{equation}

The value of $Q^2_\INT[U_{nc}^{(k)}]$ is given in \eqref{Qopt} using
$(S=2,a=1/4)$. The reduction of the error due to the use of the energy
correction term $ \left(\bracket{U^\corr} \right)^2$ can be computed
straightforwardly from eq. \eqref{Ukws}.  Finally, the contribution to
the error arising from the Madelung-Self energy
$\bracket{( \Delta U_{nc,\MS}^{(k)})^2}$ is quite involved and computationally
 intensive, and thus of little use for the purpose of tuning the
algorithm to its optimal performance point.  Nonetheless, it is shown
in Sct. \ref{sectionNumericalTest} that a reasonable estimate of the
error in the energy is obtained by dropping out the last two terms
$\bracket{ ( \Delta U_{nc,\MS}^{(k)})^2}$ and $ -\left(\bracket{U^\corr} \right)^2$ in
\eqref{summarydU} because both terms tend to cancel out mutually.  The
determination of the optimal performance point of the algorithm for
the energy can be done in just a few seconds using this last approach.
The exact expression for $\bracket{ ( \Delta U_{nc,\MS}^{(k)})^2}$ is given by
\eqref{Uself2} in App.~\ref{superappendixB}.


\section{Numerical tests}
\label{sectionNumericalTest}


In this section, the reliability of the theoretical error estimates
derived in the previous section is tested.  These theoretical
estimates will be compared to numerical errors obtained using
eq.~\eqref{def-deltaT} on configurations of a test system. The exact
numerical values $\bT^\exact(i)$ needed to use eq.~\eqref{def-deltaT}
(or eq.~\eqref{uest1} in the case of the total energy) are obtained
by a well converged standard dipolar-Ewald sum in which all quantities
are computed with a degree of accuracy $\delta \leq 10^{-10}$. The
dipolar-Ewald sum has been thoroughly tested previously against direct
sum calculations to ensure its accuracy. On the other hand, the
numerical \PPPM\ forces, torques and total electrostatic energy have
been obtained using the implementation of the dipolar \PPPM-method in
the simulation package \es~\cite{limbach06a}.  The calculations of the
error estimates have been done by truncating the aliasing sums over 
$\bm{m}=(m_x,m_y,m_z) \in \mathbb{Z}^3$ at $|m_\alpha| \leq 2 $ for $P=1$, 
and at $|m_\alpha| \leq 1 $ for assignment orders $P>1$. All the quantities in
this section are calculated using an arbitrary length unit $\cal L$
and dipole moment unit $\cal M$. Therefore, for instance, energies and
energy errors are given in units of ${\cal M}^2/{\cal L}^3$.  Hereby,
the theoretical rms error estimates will be plotted as lines,
whereas numerical rms errors will be depicted by circles.

The first test system consist of $N=100$ particles with dipole moment
of strength $\mu=1$ randomly distributed in a cubic box of length
$L=10$. Figures \ref{f1} and \ref{f2} show the rms error for forces
and torques as a function of the Ewald splitting parameter $\alpha$
for a mesh of $N_M=32$ points per direction. The real space cutoff
parameter is set to $\rcut=4$ in all plots unless specified
otherwise. From the top to the bottom, the order of the assignment
function is increased from $P=1$ to $7$.  Figure \ref{f1} shows, that
the theoretical rms error estimate (eq.~\eqref{summarydF}) gives a
good description of the numerical rms error in the whole range of
values of the Ewald splitting parameter $\alpha$.  In the inset of
figure \ref{f1}, a similar comparison is presented for different mesh
sizes. From top to bottom the number of mesh points per direction is
$N_M \in \{4, 8, 16, 32, 64\}$, and the assignment function is $P=3$. A
remarkable agreement between the theoretical error estimate and the
numerical measured error is observed.

Figure \ref{f2} shows that for torques, the rms estimates,
eq.~\eqref{summarydTau}, give also a good description of the numerical
rms error for torques in the whole range of $\alpha$'s. The inset in
figure \ref{f2} shows that if the \madelungself\ contribution is not
included in the error estimate for the torques, eq.~\eqref{dtw2}, then
large mismatches are observed at large $\alpha$'s. Nonetheless, it
should be noted, that the optimal performance point can be roughly
located even when the fluctuating errors in the \madelungself\ torques
are neglected. This behavior was confirmed for all cases studied in
this work. Thus, skipping the time consuming evaluation of the
\madelungself\ contribution (eq.~\eqref{dtw2}) is a fast and
reasonably accurate way to determine the optimal performance point for
the torques.

For the forces and torques, even the numerically computed estimate of
the rms error of a single configuration is an average over the
different dipoles (see eq.~\eqref{def-deltaT}). However, for the rms
error of the total energy \eqref{uest1}, it is is a single value.  To
obtain useful statistics, it is therefore necessary to average over a
set of configurations.

Figure \ref{f3} shows a comparison of the rms error for the energy as
a function of the Ewald splitting parameter for a mesh of $N_M=32$
points per direction.  The agreement between the theoretical and the
numerical rms errors is remarkable. The inset plot in figure \ref{f3}
shows that substantial errors arise when the energy correction term
(eq.~\eqref{Ukws}) is not taken into account (dashed lines).  The
improvement brought by the correction term decreases when the mesh
size $N_M$ is increased (at fixed number of particles $N$). Similarly
to the case of torques, a fast, though approximate, error estimate for
the energy can be obtained by dropping out the \madelungself\ and the
correction term contributions in equation \eqref{summarydU}, \ie\
\begin{equation}
  \left( \Delta U \right)^2 \approx \big(\Delta U^{(r)}\big)^2 +
  2 M^4 Q^2_\INT[U_{nc}^{(k)}].
  \label{Uapproximacio}
\end{equation}
This approach predicts quite reasonable errors (compare solid and
dashed lines in figure \ref{f4}) and has the big advantage of being
several orders of magnitude faster than the full exact error given in
\eqref{summarydU}.  It works reasonably well because it turns out that
the \madelungself\ error term $\bracket{ ( \Delta U_{nc,\MS}^{(k)})^2}$ for the energy
is quite close to the correction error term $\left(\bracket{U^\corr}
\right)^2$ and therefore they almost cancel out completely in
\eqref{summarydU}.  Therefore, it is suggested to use
\eqref{Uapproximacio} in place of \eqref{summarydU} to roughly
localize the optimal performance point of the algorithm for the
energies.

In addition, figure \ref{f4} shows that the theoretical estimates
capture the correct dependence of the rms error on the number of
particles $N$ and their dipole moments $|\bmu|$.  Various number of
particles and dipole moments were considered: $(N=1000,|\bmu|=1)$,
$(N=2000,|\bmu|=5)$, and $(N=4000,|\bmu|=25)$.


The behaviour of the error estimates for the forces, torques, and
energy in the previous figures shows that the optimal performance
point of torques and energy occur roughly at the same value of the
Ewald splitting parameter~$\alpha$.  Notice that when the parameters
of the algorithm are fixed, the highest accuracy is usually obtained
for torques, followed by the forces and the least accurate calculation
corresponds to the energy.  The optimal performance point for forces
is usually shifted slightly to higher values of the Ewald splitting
parameter $\alpha$ with respect to the optimal performance point for
torques and the energy. The shift increases with the number of mesh
points $N_M$ and the assignment order $P$. Far from the optimal point,
the behaviour of the three error estimates is, as expected, quite
different. The fact that the optimal point of energies is quite
similar to the optimal point for torques, which in turn is also not
very far from the optimal performance point for forces can be used to
do a very fast tuning of the algorithm for the three quantities:
first, the optimal performance point for forces is located using the
RMS theoretical estimate for forces (which is an immediate
calculus). In a second step, this optimal point is used as a starting
point to seek the optimal performance point for torques. In the third
stage, the optimal rms error associated to the energy can then be
straightforwardly evaluated using the error formulas for the energy
\eqref{Uapproximacio} looking in the neighbourhood of the the optimal
performance point $\alpha$ obtained for torques.

The strongest simplification done to derive the theoretical estimates
is the assumption that dipole particles are uncorrelated. Nonetheless,
tests were performed that have shown that the theoretical error
estimates are very robust against particle correlations.  In
figure~\ref{f5} the performance of the theoretical estimates is tested
for systems in which strong correlations exists among the particles. A
comparison of the theoretical rms estimates for random conformations
to the numerical rms errors obtained for forces and torques in a
typical ferrofluid simulation~\cite{wang02a} of $1000$ particles with a
diameter $\sigma \approx 1.58$ is performed. The dipolar interaction
between particles is characterized by a dipolar coupling parameter
$\lambda =3$, and a volume fraction $\phi_v = 0.3$ [which roughly
corresponds to box size $L=19$, and $M^2 \sim 11858$]. To add an extra
degree of correlation among particles, the system is under the
influence of an external magnetic field along z axis characterized by
a Langevin parameter $\alpha_L = 2$, \ie{} the characteristic energy
induced by the magnetic field is twice the thermal energy. This system
exhibits dipolar chaining, and hence a high degree of anisotropy.
Figure \ref{f5} shows that, even for this highly correlated system,
the measured errors (\PPPM\ method with $N_M=32$ and $P=7$) are close
to the theoretical estimates for randomly positioned particles. The
agreement is particularly remarkable near the optimal value
of~$\alpha$. Other tests have shown similar behaviour.  Therefore, the
theoretical estimates provide a very good guidance for the location of
the optimal performance point of the algorithm in the case of
correlated systems as well.
When the theoretical rms error estimates derived for uncorrelated systems are used to 
predict errors in non random systems,  it has been observed that  the error estimates
 for dipoles perform better than the error estimates for charges.  This difference 
 could be due to the  fact that dipolar particles have rotational degrees
 of freedom which can further reduce  the  effective degree of correlation
 respect to a  similar system made of charges.

Finally, tests have shown that the optimal influence functions as
defined in eq.~\eqref{optimized_G} ($S=3$ for forces, $S=2$ for
dipolar torques and energy) can be used interchangeably with very
little impact of the accuracy of the results, especially in proximity
to the optimal value of~$\alpha$. This is due to the exponential decay
of the reciprocal interaction $\FTnp{\phi}(\bk)$ (see
eq.~\eqref{phik}), which renders all terms $\bm{m}\neq 0$ negligible
in the numerator of eq.~\eqref{optimized_G}. Hence, in the tested cases, 
the dipolar influence functions are given in good approximation by
\begin{equation}
\label{GsimpleForm}
\FFTs{G}(\bk) = \FTnp{\phi}(\bk) \frac{\FTnp{U}^2(\bk)}{
\left( \sum_{{\bm m \in \mathbb{Z}^3}} \FTnp{U}^2(\bk+{\bm m} \frac{2\pi}{h})
\right) ^2},
\end{equation} 
which is actually the optimal lattice Green function for computing the
Coulomb energy~\cite{ballenegger08a}. The latter function has a broad
applicability because it incorporates the main effect of the \PPPM\
optimization, which is to reduce the (continuous) reciprocal
interaction by some fraction, to compensate for aliasing effects that
are inherent to the mesh calculation. 



\section{Computational performance}
\label{performance}

\subsection{Comparison against dipolar-Ewald sums} 

Due to the replacement of the Fourier transforms by $\FFT$ routines,
see eq.~ \eqref{mesh-electrostatic-field} and \eqref{mapbackF}, the
\PPPM\ algorithm is not only fast but its CPU time shows a favourable
scaling with particle number. If the real space cutoff $\rcut$ is
chosen small enough, (so that the real space contribution can be
calculated in order $N$), the complete algorithm is essentially of
order $N \log(N)$ as shown in figure \ref{f6}. In this figure, a
comparison of the presented dipolar-\PPPM\ and dipolar-Ewald sum
methods at fixed level of accuracy for the dipolar force $\Delta F
=10^{-4}$ is shown.  Parameters in both methods have been chosen to
minimize computational time given the imposed accuracy, with the only
constraint that the algorithm must satisfy the minimum image
convention ($\rcut < L/2$). Figure \ref{f6} and additional tests
performed at $\Delta F = 10^{-6}$ point out that the dipolar-\PPPM\
algorithm is faster than the dipolar Ewald sum for $N \geq 300 $.  The
inset in figure \ref{f6} shows the relative speed of the \PPPM\ to the
Ewald method as a function of the number of particles in the system.

\subsection{Constant Pressure dipolar-\PPPM\ simulations}
\label{NPTsim}

	The \PPPM\ method relies on the use of the influence function
 $\FFTs{G}(\bk)$ which depends on the box parameters,   $L$ in our cubic
 geometry.  This means that  in ensembles where the volume is not a fixed 
 quantity  the recalculation of the influence functions is needed whenever $L$ is changed. 
 The repetitive update of $\FFTs{G}(\bk)$  via 
 eq. \ref{optimized_G} or  eq. \ref{GsimpleForm} can be computationally
 expensive.  In the case of Coulomb systems, the use of \PPPM\ algorithms for 
 constant pressure simulations has been studied by H\"unenberger
 \cite{hunenberger02a} for both isotropic and anisotropic coordinate
 scalings.  The closest approach in our case to the method proposed in
 \cite{hunenberger02a} for the isotropic scaling  from a system  with
 size  $L_{(1)}$ to a system with size  $L_{(2)}$ would consist on using
 the transformations
 \begin{eqnarray}
 \alpha_{(1)} L_{(1)} =\alpha_{(2)}  L_{(2)}, \label{equalphaL}\\
 \alpha_{(1)} r_{cut,(1)} =\alpha_{(2)}  r_{cut,(2)}. \label{equalpharcut}
 \end{eqnarray}
 Indeed, due to the equality  given in eq. \ref{equalphaL} the following
 simple relation between optimal influence functions is obeyed
 \begin{equation}
 \FFTs{G}_{(1)}= \left( \frac{L_{(1)}}{L_{(2)}} \right)^2 \FFTs{G}_{(2)}
 \label{Gsrelation}
 \end{equation}
 if the mesh-size $N_M$ and influence order $P$ are unaltered.
 Under such conditions, it is simple to show from eq. \ref{Qopt} that 
 the condition \ref{equalphaL} ensures that if $(\alpha_{(1)},L_{(1)})$
 minimize   $Q^2_{int}[T^{(k)}]$ also does $(\alpha_{(2)},L_{(2)})$,
 where the relation between the value of both minimums is
 \begin{equation}
 Q^2_{int}[T^{(k)}]_{(1)}= \left( \frac{L_{(1)}}{L_{(2)}}
 \right)^{-(2S+2)} Q^2_{int}[T^{(k)}]_{(2)}.
 \label{relaQs}
 \end{equation}
It can be analogously shown that the equality given in eq. \ref{equalpharcut} leads 
to a similar scaling for the real space errors.  Thus, recalling the expressions for 
the rms error estimates (eqs. \ref{summarydF},
\ref{summarydTau}, and \ref{summarydU}),  the relation between
 the total errors of both systems is
\begin{equation}
\frac{ \Delta T_{(1)}  }{ \Delta T_{(2)}  } = \left(
\frac{L_{(2)}}{L_{(1)}} \right)^b
\label{relaAlls}
 \end{equation}
where $b=4$ for the forces, and $b=3$ for torques and energies. 
 
Therefore this approach keeps the level of accuracy set initially when we 
increase the size of the system, $L_{(1)} < L_{(2)}$ . There is however one caveat: 
if the size increases too much, it can happen that the set of parameters 
obtained from the previous scaling rules $[$ same $N_M$, same $P$, $\alpha$
and $r_{cut}$  deduced from eqs. \ref{equalphaL} and \ref{equalpharcut}
$]$ may not correspond anymore to the optimal 
point of operation of the algorithm. A practical method for dealing with 
constant (isotropic) pressure simulations is then the following: via the 
analytical error estimates determine the optimal values of the parameters for 
the smallest box-size one expects to have to simulate $L_{min}$ , use
eqs.  \ref{equalphaL} and \ref{equalpharcut}  to obtain the $\alpha$ and $r_{cut}$ for the current size L of the
system, as well as eq. \ref{Gsrelation}  to transform from the influence function calculated 
for $L_{min}$ to the one needed for $L$. If $L < L_{min}$ recompute the
influence function  via eq. \ref{optimized_G}  or eq.  \ref{GsimpleForm}. If $L \gg L_{min}$, use the error estimates to check if the 
current algorithm parameters ( $N_M$ ,$P$ and $r_{cut}$ ) are still the most optimal 
ones for speed purposes and the selected level of accuracy.
 
Unfortunately, in the case of anisotropic coordinate scalings an approach for dipoles 
similar to   the one suggested by H\"unenberger \cite{hunenberger02a} can
 be as costly as evaluating again the whole influence function. No fast alternative to
 the recalculation of the whole influence function seems to exist for this case.

\subsection{Dipoles versus charge-based system representations}
\label{compamodels}

The most simple approach for producing dipoles would be to use a pair
of opposite charges, separated by some small distance. This would be
simple, and one could use all the existing methods for simulating pure
Coulomb systems. It is therefore desirable to provide guidance about
the practical usefulness for Molecular Dynamics simulations of models
and algorithms based in true point dipole representations, as for
instance the dipolar-\PPPM\ presented in this work.

In this section we compare two different models that are intended to
represent the same physical system (a ferrofluid):  a set of $N$ particles embedded
into a cubic box of volume $V$ that interact via dipole-dipole
interaction (periodic boundary conditions used) plus a repulsive soft-core
repulsion (Weeks-Chandler-Andersen potential\cite{weeks71x}) which 
it is of the other of $k_BT$ when the distance between centers is equal
to one diameter $\sigma$. 

The model relying on true point dipoles\cite{kantorovich08x,cerda08x} 
uses a Langevin thermostat for both translational and rotational degrees of 
freedom of the particles, and the dipolar-\PPPM\ ($i \bk$-differentiation) 
algorithm  is used to  account for the long-range interactions. The  dipole moments 
have been set to $\mu=1$, and $k_BT=1$.

For the charge-based model, we have taken the most simplistic approach 
for MD simulations: the dipole is mimicked  via two point charges 
$+q$ and $-q$ which are separated by
 a distance $d$ such that $p= |q|d=\mu$
 (Gaussian units).  The movement of the two charges inside the particle 
 is constrained by a FENE potential between the charges and the
 center of the particle to force the charges to move with the particle,
 plus  a WCA and an angular  potential  acting between both charges
 in order to stabilize the dipole:
 \begin{eqnarray}
 V_{FENE}(r_{qc})&=& \frac{-K_f~r_{max}^2}{2}~~ln
 \left(1- \left( \frac{r_{qc}}{r_{max}} \right)^2 \right)\\
 V(\theta)&=&\frac{K_a}{2}(\theta-\theta_o)^2, \\
V_{WCA}(r_{qq}) &=& \left\{ \begin{array}{ll} 
     4 \epsilon \left(  \left( \frac{r_{qq}}{d} \right)^{12} - 
 \left(   \frac{r_{qq}}{d} \right)^{6} +  \frac{1}{4} 
 \right)           , &    \mbox{for  $ r_{qq} < 2^{1/6} d$ }  \\ 
                        0, &    \mbox{for  $ r_{qq} \geq  2^{1/6} d$} 
\end{array} \right. , 
\end{eqnarray}
where $r_{qc}$ is the distance of a charge to the center of the particle,
  $r_{qq}$ is the distance between both charges, and $\theta$ the angle (in radians) 
  formed by the the two charges and the center of the particle. The chosen 
 parameters for the three potentials are $r_{max}=0.8d$, $k_f=2000~k_BT$, 
 $K_a=1000~k_BT$, $\theta_o=\pi$, $\epsilon=1000~k_BT$ .
 The same Langevin thermostat  for the dipole-based model is used for
 the charge-model, but without   rotational degrees of freedom.  In this
 case, the  long-range interactions are computed using the
 Coulomb-\PPPM\ method ($i\bk$-differentiation)\cite{deserno98a,deserno98b,wang01a}. 
 
Both models have been simulated via the simulation package
 \es\cite{limbach06a,espressoweb}, which uses a velocity Verlet
 integrator. The parameters of the Coulomb and dipolar \PPPM\ algorithms have been 
  tuned in each case to the optimal values to yield  maximum speed 
 for  a force accuracy $\Delta F=10^{-4}$. Figure \ref{f7} shows the
 relative speed of the dipole-based method respect the charge-based model
  as a function of the number N of particles in the system.  The relative
speed has been computed by measuring the times $t_\mu$ and $t_q$ that the
dipole and the charge models, respectively, 
need to integrate $20000$ time steps. For the charge-model two different
separations between charges $d$ have been sampled  because the optimal value of the
Coulomb-\PPPM\ parameters ($N_M$,$P$,$r_{cut}$,$\alpha$) are observed
to depend on $d$. In general, the smaller $d$, the lengthier the calculation of the long-range forces in the charge-based model.  
The case $d=\sigma/2$ has been chosen because it represents
the limiting case for mimicking dipoles. For $d>\sigma/2$ the distance
between two charges belonging to a same particle can be larger than the
distance between  charges belonging to  different particles, and
thus the charge-model should be expected to be a poor approach to
the dipolar interaction . The  case $d=\sigma/10$ represents a more likely value of $d$.   The comparison in
figure \ref{f7} shows that the dipole-based model shows in general 
a better performance than the charge-based model for both $d=\sigma/2$
 and $d=\sigma/10$. The relative performance of the dipole-model is observed to
 increase with the reduction of the distance between charges $d$.  
 The advantage of the dipole-based model respect to the charge-based model 
 under the constrain that both models should deliver the same force
 accuracy $\Delta F=10^{-4}$ must be related to the fact that the  time needed to 
 compute several extra $FFT's$ required by the dipole-based model  plus the
 handling of the dipole rotations is in general smaller than the extra time needed
 by the charge-model  to deal with $2N$ electrostatic centers as well as
 the constrained movement  of the charges inside the particle.

Finally, it should be remarked that the time step $dt$ needed to  
run adequately the MD simulations for the charge-based
model has been found to be around two orders of magnitude smaller than
for the dipole-based model when $d=\sigma/10$, while similar time steps are
possible for  $d=\sigma/2$.  In principle this implies that for
realistic charge-based models mimicking dipoles, $d \ll \sigma/2$, 
 extra steps are needed to span  the same physical time.  Nonetheless, this difference in 
the values of the time steps could be due to the type of charge-model used
in the current comparison.  A test of the performance of
the dipolar-\PPPM\ algorithm with all possible charge-based models  is
not possible, but the present comparison illustrates 
that dipole-based models are reliable tools for simulating dipolar
systems.


\section{Conclusions}
\label{conclusions}

In this work, an extension of the \PPPM\ method of Hockney-Eastwood to
the case of dipolar interactions is presented, using the i$\bk$
differentiation scheme. This variant is expected to be the most
accurate particle-mesh based algorithm.  Optimal influence functions
that minimize the errors for dipolar forces, torques and energy have
been derived. We have shown that Madelung and self interaction terms
will arise in \emph{any} particle mesh method. We have derived
estimates of these \madelungself\ terms for the energy, force, and
torques, and proved that, for the i$\bk$-differentiation scheme, the
force \madelungself\ term is zero while the other terms are not. These 
\madelungself\ interactions are responsible for a bias in the p3m energy, 
which we suppressed by shifting the energies appropriately. Using
these results we derived accurate rms error estimates for the energy,
forces, and torques. The validity of these estimates is demonstrated
numerically by computing the errors for test systems with our \PPPM\
implementation, using various parameter sets, and comparing them to
our analytical estimates. We have further demonstrated that using our
simplified error formulas, the optimal $\alpha$ for any parameter
combinations $(N_M,\rcut,P)$ can be accurately found. Consequently,
these formulas enable to determine the parameter combination that
yields the optimal performance for any specified accuracy. This can be
conveniently done prior to running an actual simulation.

Although the derivation of the rms error assumed uncorrelated
positions and orientations of the dipoles, we numerically showed that
our estimates are sufficiently accurate also for highly correlated
systems.

The timing comparison between our dipolar-\PPPM\ algorithm and the
standard dipolar Ewald sum shows that the performance of the \PPPM\ is
superior to the standard Ewald method in systems consisting of more
than $300$ dipoles, and we see the expected (almost) linear scaling
for large particle numbers.  A protocol to speed up dipolar-\PPPM\
calculations for constant pressure simulations is 
presented in Sec \ref{NPTsim}.  In addition, the test comparing a
dipole-based model with a charge-based model to mimic simple ferrofluid
systems shows that the use of dipole-based models can be advantageous.

The somewhat tedious calculations necessary to derive our results have
been collected in the appendices for the interested reader.


\section*{Acknowledgments}

We thank E. Reznikov for help during the first stages of the present
work.  J.J. Cerd\`{a} wants to thank the financial support of Spanish
\emph{Ministerio de Educaci\'on y Ciencia}, post-doctoral grant No. EXP2006-0931,
and C. Holm acknowledges support by the DFG grant HO 1108/12-1 and the
TR6. All authors are grateful to the DAAD organization and the French
\emph{Minist\`ere des affaires \'etrang\`eres et europ\'eennes} for
providing financial support.


\appendix
\section{Building up the \PPPM\ dipolar algorithm}
\label{superappendixA}

\subsection{The optimal influence function}
\label{appendixA}
In this appendix the analytical expressions for the
optimal influence functions $\FFTs{G}$ are derived (see
eq.~\eqref{optimized_G}), and the measure $Q_\INT$ of the error for forces,
torques, and the energy is provided (see eq.~ \eqref{HE-error-measure}).
The derivation is done in close analogy to the derivation for the Coulomb
case by Hockney-Eastwood \cite{hockney81a}.

The Parseval theorem for Fourier series
\begin{equation}
  \int_V \left|  f(\br) \right|^2 \dd \br = \frac{1}{V} \SUMOKwithkzero \left| \FSp{f}(\bk) \right|^2,
\end{equation}
allows to rewrite the measure of the error $Q^2[\bT^{(k)}]$,
eq.~\eqref{HE-error-measure}, for a system containing two dipolar unit
particles $(\br_1,\hat{\bmu}_1)$ and $({\bm r}_2,\hat{\bmu}_2)$ as
\begin{eqnarray}
  Q^2_\INT[\bT^{(k)}] & = & \frac{1}{h^3~ (4\pi)^2 ~V^2}  \SUMOK \int_{h^3} d\bm{r_1} 
  \int_{{\Omega}_1} d\bm{{\Omega}_1}  \int_{{\Omega}_2} d\bm{{\Omega}_2} \left[
    |\FSp{\bT}^{(k)}(\bm{r_1},\bk,\hat{\bmu}_1,\hat{\bmu}_2 )|^2+ \right. \nonumber \\
  && \left.   
    |\FSp{\bT}^{(k,ex)}(\bk,\hat{\bmu}_1,\hat{\bmu}_2)|^2
    -2 \FSp{\bT}^{(k)}(\bm{r_1},\bk,\hat{\bmu}_1,\hat{\bmu}_2) \dotprod 
    \left[ \FSp{\bT}^{(k,ex)}(\bk,\hat{\bmu}_1,\hat{\bmu}_2) \right]^{\star}  \right]
\label{Qdevelop1}
\end{eqnarray}
where we recall that function $\bT^{(k, ex)}(\bm{r_1},\bm{r_2},\hat{\bmu}_1,
\hat{\bmu}_2)=\bT^{(k, ex)}(\bm{r_2}-\bm{r_1},\hat{\bmu}_1,\hat{\bmu}_2)$ is 
the (reciprocal) dipolar Ewald interaction between two unit dipoles (this 
interaction corresponds to the dipolar interaction of dipole 2 with dipole 1 
and with all the periodic images of dipole 1), and that $\bT^{(k)}(\bm{r_1},
\bm{r_2},\hat{\bmu}_1,\hat{\bmu}_2)$ is the corresponding interaction as 
computed with the \PPPM\ algorithm. Eq. (A.2) involves the Fourier transforms 
of these functions over $\br_2$, at fixed position $\br_1$. The Fourier 
transform of the p3m interaction  $\FSp{\bT}^{(k)}(\bm{r_1},\bk,\hat{\bmu}_1,
\hat{\bmu}_2)$ depends on the position of dipole 1 within a mesh cell, while 
the Fourier transform of the exact interaction is independent of $\br_1$ 
because of translational invariance.

The functions $\FSp{\bT}^{(k)}$ are linked to the mesh based functions 
$\FFTs{\bT}^{(k)}_M \equiv \FFT[\bT^{(k)}_M]$ by the simple relation
\begin{equation}
\FSp{\bT}^{(k)}(\bk) =  \FFTs{\bT}_M^{(k)}(\bk) ~\FTnp{U}(\bk),
\label{simple1}
\end{equation}   
which is proved below in Sct. A.2. 

In turn, $\FFTs{\bT}^{(k)}_M $ can be calculated from
eqs. \eqref{KenergiaBiased}, \eqref{mesh-electrostatic-field}, \eqref{mapbackF} and the fact that the Fast Fourier Transform of the mesh-density eq.~\eqref{mesh_density} for a single
particle system $(\br_1,\bmu_1)$ is (see ref.  \cite{ballenegger08a})
\begin{equation}
\FFTs{\brho}_M(\bk) \equiv \FFT[\brho_M(\br_m)] = \frac{1}{h^3} 
 \sum_{\bn \in \mathbb{Z}^3} \bmu_1 ~\FTnp{W}\left( \bmeta_\bn \right) ~
                  e^{-i \bmeta_\bn \dotprod \br_1 },
\label{density1p}		  
\end{equation}
where $\bmeta_\bn \equiv \bk + \frac{2\pi}{h}\bn$. Thus, for the present \PPPM\ algorithm the functions $\FSp{\bT}^{(k)}$
are
 \begin{eqnarray}
   \FSp{\bm{E}}(\bm{r_1},\bk,\bmu_1) & = &-\FFTs{\bm{D}}(\bk)~
   \FSp{\phi}_{p3m}(\bm{r_1},\bk,\bmu_1) ~  , \\
   \FSp{\bm{F}}(\bm{r_1},\bk,\bmu_1,\bmu_2) & = &
   -\FFTs{\bm{D}}(\bk) ~ \left(   {\bmu}_2  \cdot \FSp{\bm{E}}(\bk,\bmu_1)    \right),   \label{F-exp-cont}\\
   \FSp{\btau}(\bm{r_1},\bk,\bmu_1,\bmu_2)& = &(-\FFTs{\bm{D}} \left( \bk)
     \times  {\bmu}_2 \right) ~\FSp{\phi}_{p3m}(\bm{r_1},\bk,\bmu_1),    \label{tau-exp-cont}\\
   \FSp{U}_d(\bm{r_1},\bk,\bmu_1,\bmu_2) & = & 
   \left( -\FFTs{\bm{D}} \left( \bk \right)\cdot
     {\bmu}_2 \right) ~ \FSp{\phi}_{p3m}(\bm{r_1},\bk,\bmu_1),   \label{u1} 
\end{eqnarray}
where 
 \begin{eqnarray}
   \FSp{\phi}_{p3m}(\bm{r_1},\bk,\bmu_1) & = &\FTnp{U}(\bk) ~\FFTs{G}(\bk) ~
   \left(-\FFTs{\bm{D}}(\bk)\cdot
     {\bmu}_1 \right) 
      \sum_{\bmm \in \mathbb{Z}^3}  \FTnp{U}(\bmeta_{\bmm}) ~   e^{-i \bmeta_{\bmm} \dotprod \br_1 }
      ~  ,  
       \label{poti1}
 \end{eqnarray}
 and   $\FTnp{U}(\bk) \equiv \FTnp{W}(\bk)/h^3$,  $\bmeta_{\bmm} \equiv
 \bk-\frac{2 \pi }{h}\bmm $, with $\FFTs{\bm{D}} (\bk)$  defined in \eqref{definingD}.
 The quantity $\FSp{\phi}_{p3m}(\bm{r_1},\bk,\bmu_1)$ is the Fourier
 transform (over $\bm{r_2}$) of the 
electrostatic potential created at $\bm{r_2}$ by a dipole $\bmu_1$ at
$\bm{r_1}$ according to the \PPPM\ algorithm. Because of the presence of the mesh, 
that potential is not translationally invariant and depends on the position of $\bm{r_1}$ 
relative to the  mesh.

Once the functions $\FSp{\bT}$ are known, the next step involves the
calculus of the exact functions $\FSp{{\bm T}}^{(ex)}$ for the same
system. It is straightforward to show that in the case of a system
containing two particles the exact functions are
\begin{eqnarray}
  \FSp{\bm{F}}^{(ex)}(\bk,\bmu_1,\bmu_2)& = &(i\bk \cdot
  \bm{{\mu}}_2) ~  \left(i \bk \dotprod
    {\bmu}_1  \right) ~ i\bk 
  ~ \FTnp{\phi}(\bk), \\
  \FSp{\btau}^{(ex)}(\bk,\bmu_1,\bmu_2) & = &( {\bmu}_2  \times
  i \bk ) ~  \left( i \bk \dotprod {\bmu}_1 \right) 
  \FTnp{\phi}(\bk), \\
  \FSp{U}_d^{(ex)}(\bk,\bmu_1,\bmu_2) & = & -(i \bk\cdot
  {\bmu}_1) ~  \left(
    i \bk\dotprod {\bmu}_2 \right) 
  ~ \FTnp{\phi}(\bk),
\end{eqnarray}
where $\FTnp{\phi}(\bk)$ is defined in \eqref{phik}.
 In  exact calculations, as one would expect, 
only the relative distance between both particles ($\bk$ coordinate in
the reciprocal space) is relevant. 

Once the values of $\FSp{\bT}$, and $\FSp{\bT}^{(ex)}$ are known, it
is possible to simplify the expression \eqref{Qdevelop1} and arrive at
the following expression for the rms error of the reciprocal-space
components
\begin{eqnarray}
  Q_\INT^2[\bT^{(k)}]  & = & \frac{a }{9 V^2 } 
  \sum_{\bk \in \FFTs{\mathbb{M}}^3 \atop \bk \ne 0 } d\bk  \left[
    \FFTs{G}^2(\bk) ~ |\FFTs{\bm{D}}(\bk)|^{2S}  ~ \left( \sum_{m \in \mathbb{Z}^3}
      \FTnp{U}^2(\bmeta_{\bmm}) \right) \right. \nonumber \\
  && \left. + \sum_{m \in \mathbb{Z}^3} |\bmeta_{\bmm}|^{2S} ~ \left( \FTnp{\phi}(\bmeta_{\bmm}) \right)^2 \right. \nonumber \\
  && \left. -2 \FFTs{G}(\bk)\sum_{m \in \mathbb{Z}^3} 
    \left( i \bmeta_{\bmm} \dotprod \FFTs{\bm{D}}(\bk) \right)^{S} ~
    \FTnp{U}^2(\bmeta_{\bmm})~  \FTnp{\phi}(\bmeta_{\bmm})
  \right].
\label{Qdevelop3}
\end{eqnarray}
The set of parameters $(S=3,a=1)$ leads to the measure of the error in
forces, $(S=2,a=2)$ corresponds to the case of torques, and
$(S=2,a=1/4)$ must be used for the dipolar energy.  In the case of the
dipolar electrostatic field $\bm{E}$, the values of the parameters are
$(S=2, a=3)$.

The optimal influence functions for the different dipolar quantities
(force, torque, and energy) can be now obtained by minimizing
eq.~\eqref{Qdevelop3} with respect to $\FFTs{G}$,
\begin{equation}
\left. \frac{{\delta}Q_\INT^2[\bT] }{{\delta}\FFTs{G}} \right|_{\FFTs{G}_{opt}} = 0
\label{optimize}
\end{equation}
The optimal influence function expressions obtained are summarized in
eq.~\eqref{optimized_G}. Notice that the influence function optimized
for torques is the same than for the energy, which is a consequence
that for both cases it is necessary to optimize the dipolar
electrostatic field since that the dipolar energy for a particle is
$U_d = - \bmu \dotprod \bm{E}$, and its torque is $\btau = \bmu \times
\bm{E}$.  

It should be noted that the influence functions are calculated to minimize 
only errors in p3m pair interactions, neglecting errors in \madelungself\ 
interactions. In the case of forces, no further improvement
can be expected because the \madelungself\ forces are zero, but for
torques and energies further optimisation is in principle possible.
The benefit of such a full optimization is however expected to be small in 
typical systems because of the different scaling (with respect to the number 
of particles and dipoles moments) exhibited by these two sources of errors 
(see Sct. B.3).

 
\subsection{Technical proof of eq.~\eqref{simple1}}
\label{appendixb1}

The Fourier series of a function $\bT^{(k)}(\bk)$  can be written
using  the mapping-back relation (see eqs.
 \eqref{mapbackE} and \eqref{mapbackF}) as
\begin{eqnarray}
\FSp{\bT}^{(k)}(\bk)& = &\int_{V} d\br ~
  \sum_{\br_m \in \mathbb{M}^3 \atop \mic} \bT^{(k)}_M(\br_m) ~ W(\br - \br_m) ~
  e^{ -i  \bk \cdot \br } \notag \\
  & = & \sum_{\br_m \in \mathbb{M}^3} \bT^{(k)}_M(\br_m)  ~
  \FTnp{W}(\bk) ~
  e^{ -i  \bk \cdot \br_m },
 \end{eqnarray}
where the second equality follows from a change of variable (shift
theorem) and the fact the 
W(r) decays to zero on a distance shorter than half the box length.
If we replace $\bT^{(k)}_M({\mathbf r}_m)$ by the equivalent expression
$FFT^{-1}[\FFTs{\bT}^{(k)}_M]$, we obtain
 \begin{equation}
\FSp{\bT}^{(k)}(\bk) = \frac{\FTnp{W}(\bk)}{V}
  \sum_{{\bkprima} \in \widetilde{\mathbb{M}}^3} 
  \sum_{\br_m \in \mathbb{M}^3}
  \FFTs{\bT}_M^{(k)}(\bkprima) ~
  e^{ -i  (\bk-\bkprima) \cdot \br_m }.
  \label{nothing1}
\end{equation}

In order to do a further simplification, it is necessary to rewrite the sum over the mesh 
points $\br_m$ as a continuous integral  with the help of the sampling
function $\coprod(\br)$  defined as 
\begin{equation}
\coprod(\br) \equiv \sum_{\br_m \in \mathbb{M}^3} \delta(\br - \br_m)
= \frac{1}{h^3} \sum_{\bmm \in \mathbb{Z}^3}
 e^{  -i \frac{2\pi}{h} \bmm \dotprod {\br} }.
\label{samplefunction}
\end{equation}
 Thus eq. \ref{nothing1} can be rewritten as
 \begin{equation}
 \label{noti2}
\FSp{\bT}^{(k)}(\bk) = \FTnp{U}(\bk)
  \sum_{\bkprima \in \widetilde{\mathbb{M}}^3} 
  \sum_{\bmm \in {\mathbb Z}^3}
  \FFTs{\bT}_M^{(k)}(\bkprima) 
  \frac{1}{V} \int_{V} d\br~
  e^{ -i  (\bk+\frac{2\pi}{h}\bmm -\bkprima) \cdot \br }
\end{equation}
where we used $\FTnp{U}(\bk) = \FTnp{W}(\bk)/h^3$. The integral in
\eqref{noti2} divided by the volume is equal to a Kronecker delta 
$\delta_{ \bk+\frac{2\pi}{h}\bmm ,  \bkprima} $
which allows us to obtain the result
 \begin{equation}
\FSp{\bT}^{(k)}(\bk) = \FTnp{U}(\bk)
  \sum_{\bkprima \in \widetilde{\mathbb{M}}^3} 
  \sum_{\bmm \in {\mathbb Z}^3}
  \FFTs{\bT}_M^{(k)}(\bkprima) ~
\delta_{ \bk+\frac{2\pi}{h}\bmm , \bkprima } 
\end{equation}
which leads to eq. \ref{simple1}.


\section{Derivation of the rms error estimates}
\label{superappendixB}

\subsection{Errors in pair-interactions and Madelung-Self interactions}
\label{Sct_V.A}

An important point in the calculation of the rms errors is to
recognize that the error
\begin{equation}
  \Delta \bT(i) \equiv \bT(i) - \bT^\exact(i)
\end{equation}
on quantity $\bT(i)$ (energy, force or torque of a single
particle~$i$) can be understood to arise from two distinct
contributions: the interaction of a particle $i$ with all other
particles~$j\neq i$ (including the images of particles~$j$ in the
periodic replicas of the simulation box), hereby denoted by the
subscript $\INT$, and the Madelung-Self interaction
(see~\ref{Sct_Madelung}).  Thus,
\begin{align}
\label{T_int_self}
      \bT(i) &= \bT_{\mathrm{int}} (i) + \bT_{\MS}(i), \\
\label{T_int_self^exact}
   \bT^\exact(i) &= \bT_{\mathrm{int}}^\exact(i) +
   \bT_{\MS}^\exact(i),
\end{align}
and therefore the error is 
\begin{align}
\label{Decomp_T}
\Delta \bT(i) &= \Delta \bT_\INT(i) + \Delta \bT_\MS(i)
\\ \label{DeltaT_i=int+self}
&= \sum_{j\neq i} \Delta\bT_\INT(i,j) + \Delta\bT_{\MS}(i).
\end{align}
In \eqref{DeltaT_i=int+self}, $\Delta\bT_\INT(i,j)$ is the error in
the pair interaction of particle~$i$ with particle~$j$ (including the
interactions of $i$ with the images of particle~$j\neq i$).
$\Delta\bT_{\MS}(i)$ is the error in the \madelungself\ energy, force
or torque of particle~$i$. Explicit expressions for $\bT_{\MS}(i)$ can
be found in section \ref{appendixD}. The strength of a dipolar
interaction is proportional to the product of the dipole moments of
the two particles.  Setting
\begin{gather}
\label{def_x_int}
\Delta\bT_\INT(i,j) = \mu_i \mu_j \bxi_\INT(i,j)\\
\label{def_x_self}
\Delta\bT_\MS(i) = \mu_i^2 \bxi_\MS(i),
\end{gather}
\eqref{DeltaT_i=int+self} can be rewritten as
\begin{equation}
\label{xi}
\Delta \bT(i) = \mu_i \sum_{j\neq i}\mu_j \bm{\xi}_\textrm{int}(i,j) +
\mu_i^2 \bm{\xi}_{\MS}(i).
\end{equation}
By definition, $\bm{\xi}_\textrm{int}(i,j)$ and $\bm{\xi}_{\MS}(i)$ give
the direction and magnitude of the error for two \emph{unit} dipoles
[$i$ stands for $(\br_i,\FSp{\bmu}_i)$], for pair- and \madelungself\
interactions respectively. The decomposition~\eqref{xi} of the error
into an interaction and \madelungself\ contribution is a central point
in the calculation of the rms errors, because both contributions are
uncorrelated and lead to a different scaling with respect to the
dipole
moments (see further Sct.~\ref{Sct_V.B}). 

%


\subsection{Mean \madelungself\ values of the quantities}
\label{appendixD}
In this section we prove several expressions related to the mean
values of the Madelung-Self forces, torques and energies used in section
 \ref{Sct_Madelung}.


\subsubsection{Derivation of \texorpdfstring{$\bm{F}_\MS^{(k)}
\NOSINGLE =0$ }{F(k,ms)}}
\label{appendixDf1pk}

The reciprocal contribution of the \madelungself\ force of a particle is,
\begin{equation}
  \bm{F}^{(k)}_\MS \SINGLE    =  
 \frac{1}{ V}  
\SUMOK~ e^{ i\bk \dotprod \br } ~ 
 \FFTs{\bm{F}}(\br,\bk,\bmu_1=\bmu,\bmu_2=\bmu) 
\end{equation}
which using equation \ref{F-exp-cont} reduces to
\begin{eqnarray}
\label{forf2}
&=& \frac{1}{V}  \SUMOK    ~
 \left(\FFTs{\bm{D}}(\bk)\dotprod {\bmu}
 \right)^2     
 ~(-\FFTs{\bm{D}}(\bk))~ \FTnp{U}(\bk)~ \nonumber \\
& &
  \FFTs{G}(\bk)~
  \sum_{\bmm \in \mathbb{Z}^3}
  \FTnp{U}(\bmeta_{\bmm}) ~
  e^{ -i  \left( 2\pi/h \right) \bmm \dotprod \br } =0
\end{eqnarray}
The previous sum is zero because each $\bk$ term cancels out with the
corresponding $-\bk$ term (provided the lattice that is used is
symmetric).  Madelung-Self forces vanish therefore identically.


\subsubsection{Derivation of \texorpdfstring{ $\bracket{
      \btau^{(k)}_\MS \NOSINGLE }=0 $} {<tau(k)>}}
\label{appendixDtau1pk}
 
The \madelungself\ torque for a single particle can be written as
\begin{equation}
    \label{taumadself}
  {\btau}^{(k)}_\MS \NOSINGLE     =  
 \frac{1}{ V} ~  {
\SUMOK~ e^{ i\bk \dotprod \br } ~
 \FSp{{\btau}}(\br,\bk,\bmu_1=\bmu,\bmu_2=\bmu) }   
\end{equation}
where $\FSp{\btau}$ is given by eq.~\eqref{tau-exp-cont}.  Writing
explicitly the average, the following expression is obtained
\begin{eqnarray}
  \bracket{  {\btau}^{(k)}_\MS \NOSINGLE} &=&
  \frac{1}{4\pi V^2}  \int_{V}d\br  \int_{{{\Omega}_{\mu}}} d\bm{{\Omega}_{\mu}} 
  \SUMOK  ~
  e^{ i\bk \dotprod \br } \notag \\
  & &   
  \left(\FFTs{\bm{D}}(\bk)\dotprod \bmu\right) ~ 
  \left(-\FFTs{\bm{D}}(\bk) \times \bmu \right)
  \FTnp{U}(\bk) \notag \\
  & &
  \FFTs{G}(\bk)  ~\sum_{\bmm \in \mathbb{Z}^3}
  \FTnp{U}(\bmeta_m)~
  e^{ -i  \bmeta_m \dotprod \br } =0
\end{eqnarray}
This average torque vanishes because
\begin{equation}
  \int_{{{\Omega}_{\mu}}} d\bm{{\Omega}_{\mu}} ~
  (\FFTs{\bm{D}}(\bk) \times \bmu)~ 
  (\FFTs{\bm{D}}(\bk) \dotprod \bmu) 
  = 0.
\end{equation}


\subsubsection{Calculus of  \texorpdfstring{$\bracket{
U^{(k)}_{\MS}\SINGLE}$ leading to eq.~\eqref{Umean1particle}}{U(k,ms}}
\label{ukselfsingle}

The \madelungself\ energy for a single unit dipole particle can be obtained from
eq.~\eqref{u1} by setting $\bmu_1=\bmu_2=\hat{\bmu}$, $\br_1=\br$ and
evaluating the back-Fourier transform at the point $\br_2=\br$:
 \begin{eqnarray}
 \label{energymadself}
   U^{(k)}_\MS \SINGLE     &=&
 \frac{1}{2V}  {
\SUMOK ~ e^{ i\bk \dotprod \br } ~
 \FTnp{U}_d(\br,\bk,\bmu_1=\hat{\bmu},\bmu_2=\hat{\bmu}) }  \notag \\
 & =& \frac{1}{2 V }   \SUMOK  ~
 e^{ i\bk \dotprod \br }  
 \left( \FFTs{\bm{D}}(\bk)\dotprod \hat{\bmu}
 \right)^2 ~
  \FTnp{U}(\bk) ~
  \FFTs{G}(\bk) \notag \\
  & &
  \sum_{m \in \mathbb{Z}^3}
  \FTnp{U}(\bmeta_{\bmm}) ~
  e^{ -i  \bmeta_{\bmm} \dotprod \br } 
\end{eqnarray}
where $\bmeta_{\bmm} \equiv \bk + \left( 2\pi/h \right)
\bmm$. Applying the average defined in \eqref{<>} and using the
identity
\begin{equation}
\label{delta-integral}
\frac{1}{V} \int_{V}\dd \br 
e^{ -i \br \dotprod  \left( 2\pi/h
\right) \bmm  } 
= \delta_{\bmm,\bm{0}}
\end{equation}
where $\delta$ is a Kronecker symbol, and the angular integral
\begin{equation}
\frac{1}{4\pi}\int_{{{\Omega}_{\mu}}} \dd \bm{{\Omega}_{\mu}}  ~
\left(\FFTs{\bm{D}}(\bk)\dotprod {\bmu}
\right)^2 
= \frac{1}{3} ~  \FFTs{\bm{D}}(\bk)^2 ~ {\bmu}^2
\end{equation}
lead to
\begin{equation}
  \bracket{  U^{(k)}_\MS \SINGLE  } =  
 \frac{1}{6 V}
 \SUMOK ~
 \FFTs{\bm{D}}^2(\bk) ~
 \FTnp{U}^2(\bk)  ~
 \FFTs{G}(\bk).  
\end{equation}
The functions $\FFTs{\bm{D}}(\bk)$ and $\FFTs{G}(\bk)$ are periodic
over the Brillouin cells, which allows to rewrite the mean value of
the \madelungself\ energy for a single dipole particle as
eq.~\eqref{Umean1particle}.

\subsection{Scaling of the rms errors}
\label{Sct_V.B}

In this section, the scaling of the rms error estimates for the
forces, and torques with respect to $N$ and $\{\mu_i\}$ is derived
using general arguments. The results of the present section also apply
to the error of the energy of single particles, but not directly to
the error of the total energy because it involves all possible pair
interactions and an extra correction term~\eqref{Ukws}. The error of the
total energy  will be discussed apart in
section \ref{eefe}.

First, it should be noticed that the surface terms
[eq.~\eqref{Surfenergia} and last term in eq.~\eqref{Ewald E_i}] do
not lead to any error, because they are computed exactly. Therefore
from now metallic boundary conditions ($\epsilon'=\infty$) are
assumed, and surface terms are discarded . Assuming the system to be
relatively large, eq.~ \eqref{def-deltaT} can be approximated as
\begin{equation}
\label{RMS_T}
\Delta T \simeq \sqrt{ \frac{1}{N} \sum_{i} 
  \bracket{ \left( \Delta\bT(i) \right)^2  }  },
\end{equation}
by following the line of reasoning of ref.~\cite{deserno98b}.

According to~\eqref{xi}, the error $\Delta \bT(i)$ arises from errors
in pair-interactions and error in \madelungself\ interactions.  With
the energy shift~\eqref{Ukws}, the \PPPM\ algorithm is such that the
error is zero on average ($\bracket{\Delta \bT(i)}=0$), as it
should. This implies
\begin{align}
\bracket{\bm{\xi}_{\MS}(i)} &= 0\\
\bracket{\bm{\xi}_\textrm{int}(i,j)} &= 0.
\end{align}
The stronger statement that the average error of the pair-interaction
still vanishes even if dipole~$i$ is kept \emph{fixed},
\begin{equation}
\label{<xi(i,j)>_i=0}
 \frac{1}{4\pi V}\int_{V} \dd\br_j \int\dd\bm{\Omega}_j\,
  \bm{\xi}_\textrm{int}(i,j)  = 0,
\end{equation}
holds because the angular integral clearly vanishes (the integrand is
odd in $\bmu_j$).  The property~\eqref{<xi(i,j)>_i=0} implies in
particular that
\begin{align}
\label{<x.xself>=0}
\bracket{\bxi_\INT(i,j) \dotprod \bxi_\MS(i)} &= 0 \\
\label{<xself.xself>}
\bracket{\bxi_\INT(i,j)\cdot\bxi_\INT(i,k)} &= \delta_{j,k} \bracket{\bxi_\INT^2(i,j)}.
\end{align}
The mean-square error $\bracket{ \Delta\bT(i)^2 } $ in~\eqref{RMS_T} becomes
\begin{align}
\notag
\bracket{ \Delta\bT^2(i) } 
&= \bracket{\Big(
\mu_i \sum_{j\neq i}\mu_j \bm{\xi}_\textrm{int}(i,j) + \mu_i^2
\bm{\xi}_{\MS}(i)
\Big)^2 }  \\
\notag
&= \mu_i^2
 \bracket{  \big( \sum_{j\neq i} \mu_j \bm{\xi}_\textrm{int}(i,j) \big)^2 } 
+ \mu_i^4  \bracket{ \bm{\xi}^2_{\MS}(i)  }  \\
&= \mu_i^2  \sum_{j\neq i} \mu^2_j \bracket{\bxi_\INT^2(i,j)} + \mu_i^4
 \bracket{ \bm{\xi}^2_{\MS}(i)  } .
\end{align}
where the second equality follows from~\eqref{<x.xself>=0} and the
third equality from~\eqref{<xself.xself>}.  The mean-square errors of
the pair and \madelungself\ interactions,
\begin{align}
\label{def_Q_int}
\bracket{\bxi_\INT^2(i,j)}&=Q^2_\INT[\bT] \\
\bracket{\bm{\xi}^2_{\MS}(i) }&=Q^2_\MS[\bT],
\end{align}
do not depend on the chosen pair of particles ($i,j$) by definition of
the configurational average.  The mean-square error on particle~$i$
reduces (using  $(M^2- \mu_i^2 ) \simeq M^2$) to
\begin{eqnarray}
\bracket{\Delta\bT^2(i)} &\simeq& \mu_i^2  M^2    Q^2_\INT + \mu_i^4 Q^2_\MS
\end{eqnarray}
Eventually, it is found that the rms (total) error~\eqref{RMS_T} can
be expressed as
\begin{equation}
\label{Scaling_rms}
\Delta T^2 \simeq  \frac{  M^4   Q^2_\INT + \sum_i \mu_i^4 \, Q^2_\MS}{N},
\end{equation}
where, using \eqref{def_Q_int},
\begin{equation}
\label{Q2_int_explicit}
Q^2_\INT[\bT] = \frac{1}{(4\pi)^2 h^3 V}
\int_{h^3}\dd\br_1\int_{V}\dd\br_2
\int\dd\bm{\Omega}_1\int\dd\bm{\Omega}_2 \, \bxi^2_\INT(1,2)
\end{equation}
is the mean-square error in the pair interaction between two unit
dipoles (see eq.~\eqref{def_x_int}) and
\begin{equation}
\label{Q2_self_explicit}
 Q^2_\MS[\bT] = \frac{1}{(4\pi) h^3}
\int_{h^3}\dd\br_1 \int\dd\bm{\Omega}_1
\, \bxi^2_{\MS} (1) 
\end{equation}
is the mean-square error in the \madelungself\ interaction of a unit
dipole (see eq.~\eqref{def_x_self}). Notice that the average over $\br_1$
in eqs. \eqref{Q2_int_explicit} and \eqref{Q2_self_explicit} can be
restricted to a single mesh cell $h^3$ thanks the periodicity of the
system.

The result~\eqref{Scaling_rms} exhibits the scaling of the rms error
$\Delta T$ with respect to the number of particles and the magnitudes
of the dipole moments.

It is important to stress that our result for the scaling of $\Delta
T$ takes into account not only the contributions from errors in
pair-interactions, but also errors in \madelungself\
interactions. When using standard dipolar Ewald sums, rms errors in
\madelungself\ interactions are negligible (at least if the energies
are correctly shifted~\cite{errorsmall} and \eqref{Scaling_rms}
reduces to the expression found in ref.~\cite{wang01a} for the scaling
of the error. By contrast, the errors due to \madelungself\ dipolar
interactions play an important role when using Particle-Mesh methods,
because of the loss of accuracy brought by the discretization of the
system onto a mesh.


\subsection{Explicit formulas for the rms errors}
\label{Sct_V.C}

To use the error estimate~\eqref{Scaling_rms}, we need to know the
mean-square errors $Q^2_\INT$ and $Q^2_\MS$, which measure,
respectively, errors in the pair-interaction $\bT_\INT(i,j)$ and
errors in the \madelungself\ interaction $\bT_\MS(i)$. These errors
depend on the details of the method employed to compute them (here the
\PPPM\ algorithm), but are independent of the simulated system.  In
this section explicit theoretical expressions for these errors are
derived. These expressions are functions of the ``methodological''
dimensionless parameters ($\alpha L$, $\rcut/L$, $N_M=L/h$ and
$P$). It should be recalled that surface terms are discarded by
setting metallic boundary conditions, because these terms do not play
any role in the error estimates.

The quantity $\bT$ (=  force, electrostatic field, torque or energy) is
computed as a sum of a real-space contribution $\bT^{(r)}$ and a
reciprocal-space contribution $\bT^{(k)}+\bT^\self$ [$\bT^\self$
vanishes in the case of the force, see eqs. \eqref{U_Ewald},
\eqref{Ewald F_i} and \eqref{Ewald E_i}]. If the errors in these two
contributions are assumed to be statistically independent, it can be
written with \eqref{Scaling_rms} and \eqref{Q2_int_explicit} in mind,
\begin{equation}
\label{DeltaT^k}
 \left( \Delta T \right)^2 \simeq 
 \left(\Delta T^{(r)} \right)^2 + 
 \left(\Delta T^{(k)} \right)^2
\end{equation}
where $\Delta T^{(r)}$ is the rms error arising from the real-space
contribution, and $\Delta T^{(k)}$ is the rms error arising from the
reciprocal-space contribution. These two rms errors are given by eqs.
\eqref{Scaling_rms}-\eqref{Q2_self_explicit}, in which the mean-square
errors $Q^2_\INT$ and $Q^2_\MS$ are computed with the direct-space,
respectively reciprocal-space, contribution to $\Delta\bT(i)$ only.

\subsubsection{ Error estimates for real-space contributions}

Introducing decomposition \eqref{Decomp_T}, the real-space
contribution to the rms error~\eqref{DeltaT^k} splits into two terms
\begin{equation}
\label{deo}
\big(\Delta T^{(r)}\big)^2 = \big(\Delta T^{(r)}_\INT\big)^2 +
\big(\Delta T^{(r)}_\MS\big)^2
\end{equation}
where $\Delta T^{(r)}_\INT$ is the rms error of (real) pair
interactions and $\Delta T^{(r)}_\MS$ is the rms error of (real)
Madelung interactions.  No cross-term appears in~\eqref{deo} because
of property \eqref{<x.xself>=0}.  $\big(\Delta T^{(r)}_\MS\big)^2$ is
negligible due to the fast decay of the real-space contribution. Thus,
$\big(\Delta T^{(r)}_\MS \big)^2=0$, and the real-space rms errors of
the \PPPM\ method approximately coincide with those derived for the
dipolar Ewald sum method~\cite{wang01a}, because real-space
contributions are evaluated identically in both methods. These error
estimates are given by \eqref{DeltaF^r},
\eqref{DeltaT^r} and \eqref{DeltaU^r} (see also ref. ~\cite{wang01a}).  Notice the exponential decay
$\exp(-\alpha^2 \rcut^2)$ of the error with the real-space cutoff
distance~$\rcut$.

\subsubsection{Error estimates for reciprocal-space contributions}

Introducing decomposition \eqref{Decomp_T}, the reciprocal
contribution to the rms error~\eqref{DeltaT^k} splits into two terms
\begin{equation}
\label{DeltaT=int+self}
\big(\Delta T^{(k)}\big)^2 = \big(\Delta T^{(k)}_\INT\big)^2 + \big(\Delta T^{(k)}_\MS\big)^2
\end{equation}
where $\Delta T^{(k)}_\INT$ is the rms error of (reciprocal) pair
interactions and $\Delta T^{(k)}_\MS$ is the rms error of (reciprocal)
\madelungself\ interactions. No cross-term appears
in~\eqref{DeltaT=int+self} because of property \eqref{<x.xself>=0}.
By~\eqref{Scaling_rms}, these two contributions scale like
\begin{gather}
\label{DeltaT^k_int}
\big(\Delta T^{(k)}_\INT\big)^2 = \frac{ M^4}{N} Q^2_\INT[\bT^{(k)}] \\
\label{DeltaT^{(k)}_self}
\big(\Delta T^{(k)}_\MS\big)^2 = \frac{\sum_i \mu_i^4}{N} Q^2_\MS[\bT^{(k)}]
\end{gather}
where $Q^2_\INT[T^{(k)}]$ (resp. $Q^2_\MS[T^{(k)}]$) is the
contribution to the mean-square error~\eqref{Q2_int_explicit} (resp.
\eqref{Q2_self_explicit}) associated to the reciprocal interaction
$\bT^{(k)}$.  The problem of predicting the rms errors of the \PPPM\
algorithm is now reduced to finding explicit expressions for the
functions $Q^2_\INT[T^{(k)}]$ and $Q^2_\MS[T^{(k)}]$. The detailed
calculation of these quantities is performed in
 section~\ref{appendixA} for the pair interactions, and
 section~\ref{appeE}~for the \madelungself\
interactions. For the total energy see section~\ref{eefe}.\\

\subsubsection*{(b.1) rms error in pair-interactions: $\Delta T^{(k)}_\INT$}

The lattice Green function $\FFTs{G}(\bk)$ is determined in the \PPPM\
method by the condition that it minimizes the rms error $\Delta
T^{(k)}_\INT$ of the (reciprocal) pair-interaction. The minimization
of this rms error was performed in App.~\ref{appendixA}, where is
it shown that the minimal errors are given by eq.~\eqref{Qopt} where
in the case of forces, we have to use the set of parameters
$(S=3,a=1)$, for torques $(S=2,a=2)$, and $(S=2,a=1/4)$ for the
energy.  It should be noticed that eq.~\eqref{Qopt} reduces to the rms
error corresponding to Coulomb forces when the parameters are set to
$(S=1, a=1)$ and the factor $1/9$ is dropped \cite{deserno98a}.  When
the optimal lattice Green function~\eqref{optimized_G} is used, the
(reciprocal) rms error in pair-interaction is given by inserting
\eqref{Qopt} into \eqref{DeltaT^k_int}.\\

\subsubsection*{(b.2) rms error in \madelungself\ interactions: $\Delta T^{(k)}_\MS$}
\label{appeE}

From \eqref{DeltaT^{(k)}_self} and \eqref{Q2_self_explicit}, the rms
error in \madelungself\ interactions involve the quantity
\begin{equation}
Q^2_\MS[\bT^{(k)}] = 
\bracket{    \left(   \bT_\MS^{(k)} \SINGLE -  \bT_\MS^{(k,ex)} \SINGLE  \right)^2  },
\label{Madelungdt2}
\end{equation}
where $ \bT_\MS^{(k)} \SINGLE $ is the \PPPM\ Madelung-Self
interaction defined in section \ref{Sct_Madelung} for a unit dipole. The exact
\madelungself\ interaction $\bT_\MS^{(k,ex)} \SINGLE $ is non-zero
only in the case of the energy. Since the \PPPM\ \madelungself\ force
is identically zero (see section \ref{appendixD}) the rms error
vanishes for this quantity:
\begin{equation}
\Delta F_\MS^{(k)}=0.
\end{equation}
According to \eqref{taumadself}, the rms error of \madelungself\
torques is given by
\begin{eqnarray}
\nonumber
Q^2_\MS[\btau^{(k)}] &=& \bracket{  \left( {\btau}^{(k)}_\MS \SINGLE \right)^2   }   = 
\frac{1}{4\pi V^3}
\int_{V} \dd \br  ~\int_{{{\Omega}_{\mu}}} \dd\bm{{\Omega}_{\mu}} 
\SUMOKone \SUMOKprima
\left( \FFTs{\bm{D}}(\bkone)\dotprod {\hat{\bmu}}
\right)^2 ~
\left( \FFTs{\bm{D}}(\bkprima)\dotprod
\hat{\bmu} \right)^2 \\ \nonumber
 & &
  \left[ \left(-\FFTs{\bm{D}}(\bkone) \times
  \hat{\bmu} \right) 
  \dotprod 
  \left(-\FFTs{\bm{D}}(\bkprima) \times
  \hat{\bmu} \right) 
 \right] ~
 \FTnp{U}(\bkone) \FFTs{G}(\bkone) ~
 \FTnp{U}(\bkprima) \FFTs{G}(\bkprima)\\ \nonumber
 & & 
\left( \sum_{\bmm \in \mathbb{Z}^3}   \FTnp{U}(\bkone_{\bmm}) ~ 
e^{ -i \bkone_{\bmm} \dotprod \br } \right)  
\left( \sum_{\bn \in \mathbb{Z}^3}   \FTnp{U}(\bkprima_\bn) ~ 
e^{ -i \bkprima_\bn \dotprod \br } \right)
\\ \nonumber
& &
e^{ i  \left( \bkone   +\bkprima
\right) \dotprod \br  }
\end{eqnarray}
where $\bkone_{\bmm} \equiv \bkone +  \left(
2\pi/h \right) \bmm$ and $\bkprima_\bn \equiv
\bkprima + \left( 2\pi/h \right) \bn$.
 The integral in eq.~\eqref{delta-integral}, and the angular integral 
\begin{equation}
\int_{{{\Omega}_{\mu}}} d\bm{{\Omega}_{\mu}} 
\left ( \bm{a} \dotprod {\bmu} \right)
\left( \bm{b} \dotprod {\bmu} \right)
\left[ \left( \bm{a} \times {\bmu} \right) \dotprod 
( \bm{b} \times {\bmu}) \right]
=\frac{2 \pi \bmu^4}{3} h(\bm{a},\bm{b})
\end{equation}
where $h(\bm{a},\bm{b})$ is given by eq.~\eqref{hequation}, lead to
\begin{eqnarray}
Q^2_\MS[\btau^{(k)}] & = & 
\frac{1}{6 V^2}
\SUMOKone \SUMOKprima~
 h\left(  \FFTs{\bm{D}}(\bkone)  ,  \FFTs{\bm{D}}(\bkprima)  \right) ~
 \FTnp{U}(\bkone) \FFTs{G}(\bkone) ~\\ \nonumber
 &  & \FTnp{U}(\bkprima) \FFTs{G}(\bkprima) ~
  \left( \sum_{\bmm \in \mathbb{Z}^3}   
  \FTnp{U}( \bkone_\bmm  ) ~
  \FTnp{U}( \bkprima_\bmm  ) \right)  \nonumber
\end{eqnarray}
Finally, using the fact that $\FFTs{\bm{D}}(\bk)$ and $\FFTs{G}(\bk)$
are periodic over the Brillouin cells, the mean square \madelungself\
torque for the reciprocal contribution reduces to the expression given
in \eqref{dtw2}.


\subsection{Rms error for  the total corrected energy}
\label{eefe}

A theoretical estimate can be derived for the rms error of the total
energy $\Delta U_\pppm$ in eq.~\eqref{uest1}. Hereby in order to avoid
confusions the values related to non-corrected energies will be
identified with a subindex $(nc)$.  As in the case of forces and
torques the error is split into real and space contributions
\begin{equation}
\left( \Delta U_\pppm \right)^2 = \left( \Delta U^{(r)}_\pppm \right)^2
+\left( \Delta U^{(k)}_\pppm \right)^2. 
\end{equation}
As in the case of forces and torques, the fast decay of the real-space
interaction  makes the \madelungself\  contribution arising from the
real-space negligible. Thus,  the value of $\left( \Delta
  U^{(r)}_\pppm \right)^2$ is the same than in Ewald calculations
\cite{wang01a} (see eq.~\eqref{DeltaU^r}).

The rms error of the reciprocal-part of the energy is by definition
\begin{eqnarray}
\label{Urms1}
\left( \Delta U^{(k)}_\pppm \right)^2 &:=& 
\bracket{ \left(   U^{(k)}_\pppm - U^{(k)}  \right)^2},
\end{eqnarray}
where $U^{(k)} $ is the exact reciprocal-space energy given by eq.
\eqref{U^k}.  The energy correction term \eqref{Ukws} is fully
associated to the calculations in the reciprocal-space when \mic\ is
used. Thus, eq.~\eqref{Urms1} can be rewritten in terms of
$\bracket{U^\corr}$ and the reciprocal-space error of the
non-corrected energy $\Delta U_{nc}^{(k)}$ as
\begin{eqnarray}
  \left( \Delta U^{(k)}_\pppm \right)^2 &:=& 
  \bracket{ \left( \Delta U_{nc}^{(k)} + \bracket{U^\corr}  \right)^2} \\
  &=&  \bracket{ \left( \Delta U_{nc,\INT}^{(k)} +  \Delta
      U_{nc,\MS}^{(k)}+ \bracket{U^\corr}  \right)^2}.
\end{eqnarray}
applying that
\begin{eqnarray}
  \bracket{ \Delta U_{nc,\INT}^{(k)} }&=& 0 \\
  \bracket{ \Delta U_{nc,\INT}^{(k)}  \Delta U_{nc,\MS}^{(k)} }&=& 0 
\end{eqnarray}
the rms error for the reciprocal contribution is
\begin{eqnarray}
  \left( \Delta U^{(k)}_\pppm \right)^2 &:=& 
  \bracket{ \left( \Delta U_{nc,\INT}^{(k)} \right)^2}  +  
  \bracket{ \left( \Delta U_{nc,\MS}^{(k)}+ \bracket{U^\corr}  \right)^2}.
\end{eqnarray}
If the relation $ \bracket{ \Delta U_{nc,\MS}^{(k)} } \approx -
\bracket{U^\corr} $ is used, then
\begin{eqnarray}
  \left( \Delta U^{(k)}_\pppm \right)^2 &:=& 
  \bracket{ \left( \Delta U_{nc,\INT}^{(k)} \right)^2}  +  
  \bracket{ \left( \Delta U_{nc,\MS}^{(k)}\right)^2} -  
  \left(\bracket{U^\corr}  \right)^2,
\end{eqnarray}
which shows that the correcting term $\bracket{U^\corr}$, in addition
to removing the systematic bias in the reciprocal-energies, also
reduces the fluctuating errors of the reciprocal-space self-energies
by an amount $ - \left(\bracket{U^\corr} \right)^2$.  In the following
sections (a, b, and c) it is shown that
\begin{eqnarray}
\label{Uint2}
\bracket{ \left( \Delta U_{nc,\INT}^{(k)} \right)^2} &=& 2 ~M^4~Q^2_\INT [U_{nc}^{(k)}] \\
\end{eqnarray}
and $\bracket{ \left( \Delta U_{nc,\MS}^{(k)}\right)^2}$ is given by
\begin{eqnarray}
\label{Uself2}
\bracket{ \left( \Delta U_{nc,\MS}^{(k)}\right)^2} &=& 
M^4~\left[  \left(  \bracket{ U_{\MS}\SINGLE} \right)^2 - 2 ~ 
  U^{\exact}_{\MS}\SINGLE ~  \bracket{
    U_{\MS}\SINGLE}   + \left( U^{\exact}_{\MS}\SINGLE    \right)^2
\right]  \notag \\
& &
+ \left( \sum_{i=1}^N \bmu_i^4 \right) ~ \left[    \bracket{
    \left(  U_{\MS}\SINGLE \right)^2  }    -    \left( \bracket{
      U_{\MS}\SINGLE }   \right)^2 \right]
 \end{eqnarray}
 where the mean \PPPM\ \madelungself\ energy of a unit dipole particle
 $ \bracket{ U_{\MS}\SINGLE} $ is \eqref{Umean1particle}, the exact
 \madelungself\ energy $U^{\exact}_{\MS}\SINGLE$ is \eqref{U_MS^ex}
 and $\bracket{U^\corr}$ is given by eq.~\eqref{Ukws}.
 
 On the other hand, in section c is shown that the mean square
 \madelungself\ energy of a unit dipole in the \PPPM\ calculation is
\begin{eqnarray}
  \bracket{ \left(  U_{\MS}\SINGLE \right)^2  }   &=&  \frac{1}{120 V^2}
  \sum_{\bkone  \in \FFTs{\mathbb{M}}^3 \atop \bkone \ne 0}
  \sum_{\bkprima  \in \FFTs{\mathbb{M}}^3 \atop \bkprima \ne 0 }  
  \FFTs{G}(\bkone)~
  \FFTs{G}(\bkprima)~   
  f(\FFTs{\bm{D}}(\bkone) ,\FFTs{\bm{D}}(\bkprima) )
  \nonumber \\ 
  & & 
  \sum_{\bm{t} \in \mathbb{Z}^3} \sum_{\bm{l} \in \mathbb{Z}^3} \sum_{\bmm \in \mathbb{Z}^3}    \left[
    \FTnp{U}(\bkone_{\bm{t}} ) ~ 
    \FTnp{U}(\bkprima_{\bm{l}})~
    \FTnp{U}( \bkone_{\bm{tm}}~ ) 
    \FTnp{U}( \bkprima_{\bm{lm}} )  \right]  , \label{averageUmssingle2}
\end{eqnarray}
where
\begin{eqnarray}
  \label{fequation}
  f( \bm{a}, \bm{b} ) & =&  \left(
    \frac{| \bm{a} + \bm{b } |^4+| \bm{a} - \bm{b }  |^4}{2}-\bm{a}^4 -\bm{b}^4 \right)
\end{eqnarray}
with $\bmeta_{\bm{\alpha}} \equiv \bk + (2\pi/h) \bm{\alpha}$ , and
$\bmeta_{ \bm{ \alpha \beta}} \equiv \bk +(2\pi/h)
(\bm{\alpha}+\bm{\beta})$.  Similar techniques to the ones used in the
case of torques can reduce by several orders of magnitude the
computational effort, rendering its exact calculation
feasible, although for practical purposes to determine the rms energy
error it is advisable to use the approach stated in Sct.
\ref{totalenergyerror}.


\subsubsection{Derivation of 
\texorpdfstring{$\bracket{ \left( \Delta U_{nc,\INT}^{(k)}\right)^2}
$}{<U(k,nc,int)\texttwosuperior>}} 
\label{subsectionE3}

In this section, the mean square value of the pair energy of the
non-corrected interactions is derived.  Using eq.~\eqref{xi} the pair
energy of a system of $N$ particles can be written as
\begin{equation}
 \bracket{ \left( \Delta U_{nc,\INT}^{(k)}\right)^2} = \sum_{i}^{N}
 \sum_{j}^{N} \sum_{k \ne i }^{N}\sum_{l \ne j }^{N}
  \mu_i ~ \mu_j ~\mu_k ~\mu_l  ~
  \bracket{\bm{\xi}_{\INT}(i,k) \dotprod \bm{\xi}_{\INT}(j,l)}, 
\end{equation}
applying
\begin{equation}
  \bracket{\bm{\xi}_{\INT}(i,k) \dotprod  \bm{\xi}_{\INT}(j,l)} = 
  \left(\delta_{i,j} ~\delta_{k,l}  +  \delta_{i,l} ~\delta_{k,j}  \right)
  ~ \bracket{ \bm{\xi}^2_{\INT}(i,k) }
\end{equation}
the rms error can be written (using the approach $(M^4-\sum_i
\bmu_i^4) \simeq M^4 $ as
\begin{eqnarray}
  \bracket{ \left( \Delta U_{nc,\INT}^{(k)}\right)^2} &=& 2 \sum_{i}^{N}
  \sum_{k \ne i }^{N}
  \mu_i^2 ~\mu_k^2 ~\bracket{  \bm{\xi}^2_{\INT}(i,k)  } \\
  &\approx&  2  M^4 ~\bracket{ \bm{\xi}^2_{\INT}(1,2)  } 
\end{eqnarray}
where $\bracket{  \bm{\xi}^2_{\INT}(1,2)}=Q^2_\INT[U^{(k)}_{nc}]$ (see
eq.~\eqref{Q2_int_explicit}).


\subsubsection{Derivation of \texorpdfstring{$\bracket{ \left( \Delta
        U_{nc,\MS}^{(k)}\right)^2} $}{<U(k,nc,ms)\texttwosuperior>}}
\label{subsectionE2}

In this section, the mean square value of the \madelungself\ energy of
the non-corrected interactions is derived.  For a system of $N$
particles it can be expressed in terms of the non corrected \PPPM\ and
exact \madelungself\ energies of each particle, $U_{nc,\MS}^{(k)}(i)$
and $U^{\exact}_\MS (i)$ respectively, as
\begin{equation}
 \bracket{ \left( \Delta U_{nc,\MS}^{(k)}\right)^2} = 
\bracket{ \left[  \sum_{i}^{N} \left(U_{nc,\MS}^{(k)}(i) - U^{\exact}_\MS (i) \right) \right] ~ 
\left[ \sum_{j}^{N} \left(U_{nc,\MS}^{(k)}(j) - U^{\exact}_\MS (j) \right) \right]^{*} } 
\end{equation}
 where the asterisk denotes
complex conjugate, $U^{\exact}_\MS (i)= \bmu_i^2 U^{\exact}_\MS \SINGLE$
with $U^{\exact}_\MS \SINGLE$ given in \eqref{U_MS^ex}, and
\begin{equation}
U_{nc,\MS}^{(k)}(i)  = \bmu^2_i U_{nc,\MS}^{(k)}\SINGLE  =  \frac{\bmu^2_i}{2}\FSpletters^{-1}_{\bk \ne 0}[
\FFTs{U}_d^{p3m}(\br_i,\bk,\hat{\bmu})],
\end{equation}
notice that the surface energy terms have been dropped because they
would be the same and would just cancel out. Some algebra, and a
careful separation of the terms $i \ne j$ from the $i = j$ terms,
leads to eq.~\eqref{Uself2}.

\subsubsection{ Proof of \texorpdfstring{$\bracket{  \left( U^{(k)}_\MS
\SINGLE \right)^2}$}{<(U(k,ms)\SINGLE)\texttwosuperior>}}

Taking the square of eq.~\eqref{energymadself} and using the average
given in \eqref{<>} we get
\begin{eqnarray}
\bracket{   \left( U^{(k)}_\MS \SINGLE
\right)^2   }  & = & 
\frac{1}{ 16 \pi V^3}
\int_{V}\dd\br  \int_{{{\Omega}_{\mu}}} \dd\bm{{\Omega}_{\mu}} 
\SUMOKone \SUMOKprima
  \left( \FFTs{\bm{D}}(\bkone)\dotprod \hat{\bmu}
  \right)^2 ~ \\ \nonumber
& &
 \left(\FFTs{\bm{D}}(\bkprima)\dotprod
 \hat{\bmu} \right)^2 ~
 \FTnp{U}(\bkone) \FFTs{G}(\bkone) ~
 \FTnp{U}(\bkprima) \FFTs{G}(\bkprima) ~
\left( \sum_{\bmm \in \mathbb{Z}^3}   \FTnp{U}(\bkone_{\bmm}) ~
e^{ -i \bkone_{\bmm} \dotprod \br } \right) \\
\nonumber
 & &
\left( \sum_{\bn \in \mathbb{Z}^3}  
\FTnp{U}(\bkprima_{\bn}) ~ 
e^{ -i \bkprima_{\bn} 
\dotprod \br } \right) ~
e^{ i  \left( \bkone \dotprod \br 
 + \bkprima \dotprod \br  \right)}. 
\end{eqnarray}
The integral in eq.~\eqref{delta-integral} and the angular integral
\begin{equation}
\int_{{{\Omega}_{\mu}}} d\bm{{\Omega}_{\mu}} ~
 \left( \bm{a} \dotprod {\bmu} \right)^2 ~
\left( \bm{b} \dotprod  {\bmu} \right)^2
=\frac{2 \pi \bmu^4}{15} f(\bm{a},\bm{b})
\end{equation}
 where  $f(\bm{a},\bm{b}) $ is given in eq.~ \eqref{fequation}, 
  lead to 
\begin{eqnarray}
  \bracket{  \left( U^{(k)}_\MS \SINGLE
    \right)^2   }  & = & 
  \frac{ 1}{120 V^2}
  \SUMOKone \SUMOKprima ~
  f \left( \FFTs{\bm{D}}(\bkone)  , \FFTs{\bm{D}}(\bkprima)        \right) ~
  \FTnp{U}(\bkone) \FFTs{G}(\bkone) \nonumber\\ 
  &  & \FTnp{U}(\bkprima) \FFTs{G}(\bkprima)
  \left( \sum_{\bmm \in \mathbb{Z}^3}   
    \FTnp{U}( \bkone_\bmm ) 
    \FTnp{U}( \bkprima_\bmm ) \right).
\end{eqnarray}
Finally, taking into account that $\FFTs{\bm{D}}(\bk)$ and
$\FFTs{G}(\bk)$ are periodic over the Brillouin cells, the final
expression for the rms \madelungself\ energy is
eq.~\eqref{averageUmssingle2}.


\bibliographystyle{phaip}
\bibliography{paper,comentaris}

\begin{thebibliography}{10}
\bibitem{errorsmall}
The contribution to the (total) rms error arising from errors in \madelungself\
  energies is negligible only if the Ewald energy is corrected to compensate
  systematic cutoff errors in the \madelungself\ energies of the particles.
  This correction, which is the analog of $\bracket{U^\corr}$ for standard
  Ewald sums, is termed the ``diagonal'' correction in refs.
  \cite{kolafa92a,wang01a}.
\end{thebibliography}


\clearpage
\section{CAPTION LIST,  TABLES AND FIGURES}

\begin{figure} 
  \centering
  \includegraphics[width=\linewidth,clip=true]{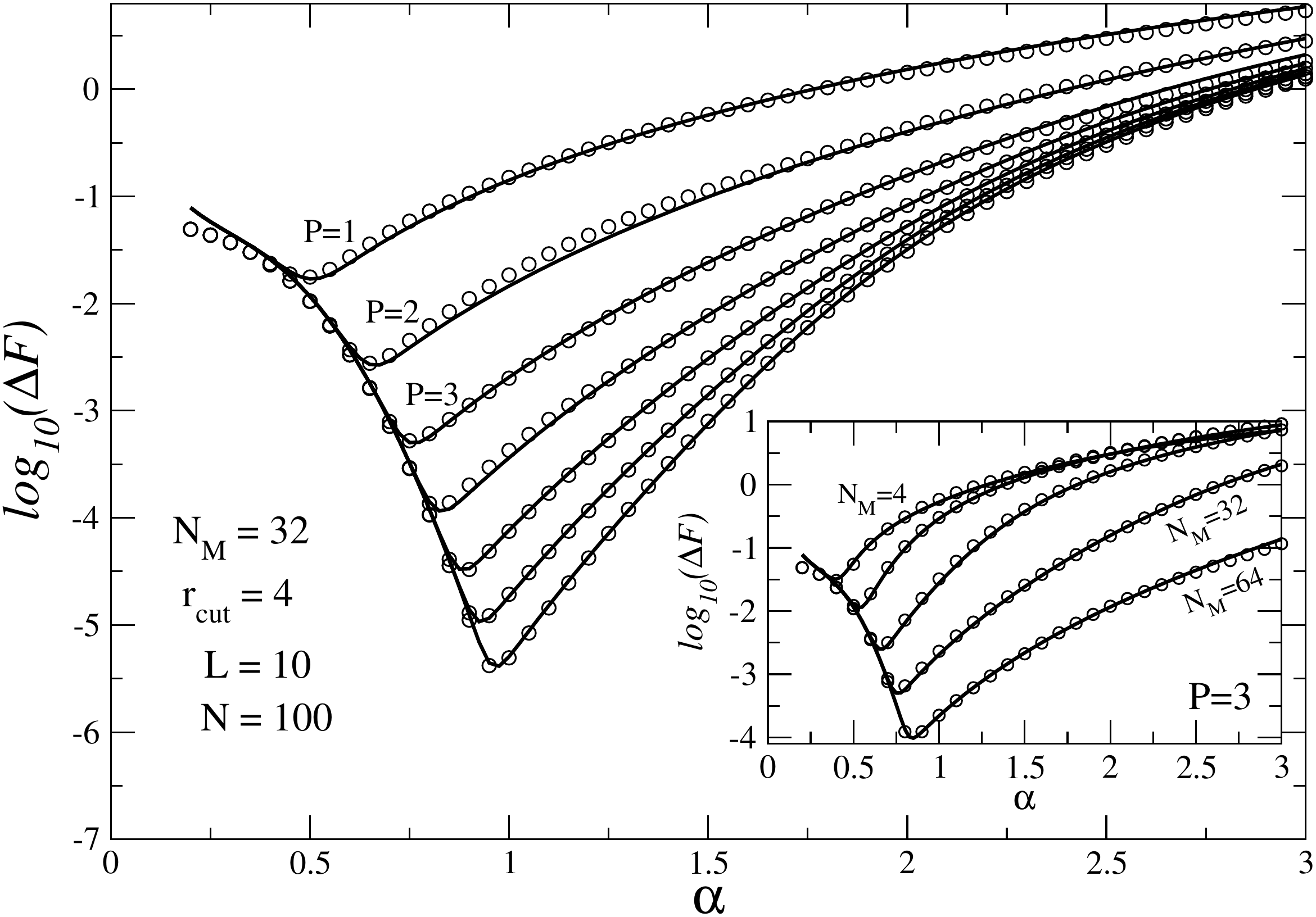}
  \caption{Figure 1}
  \label{f1}
\end{figure}

\begin{figure}
  \centering
  \includegraphics[width=\linewidth,clip=true]{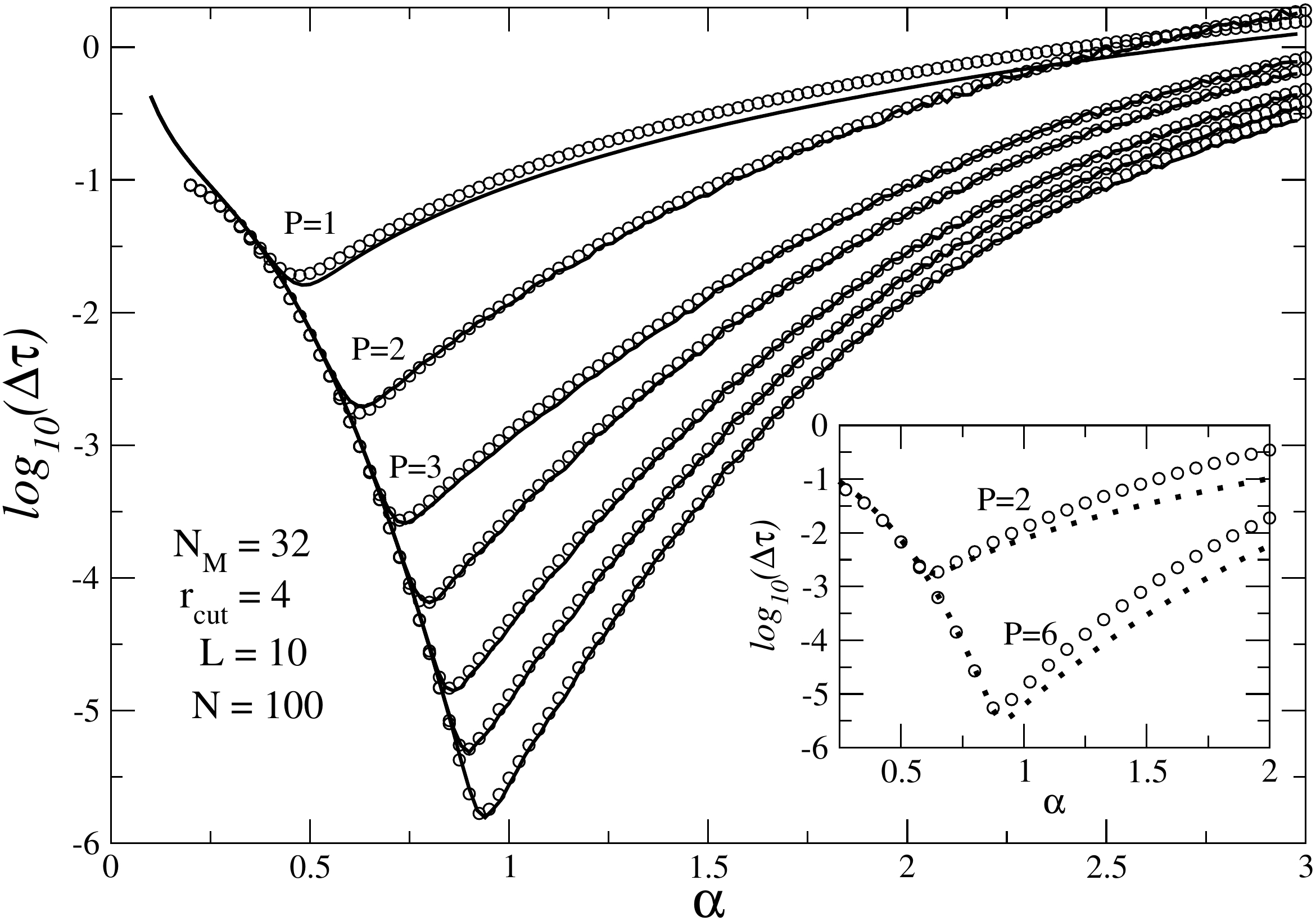}
  \caption{Figure 2}
  \label{f2}
\end{figure}

\begin{figure}
  \centering
  \includegraphics[width=\linewidth,clip=true]{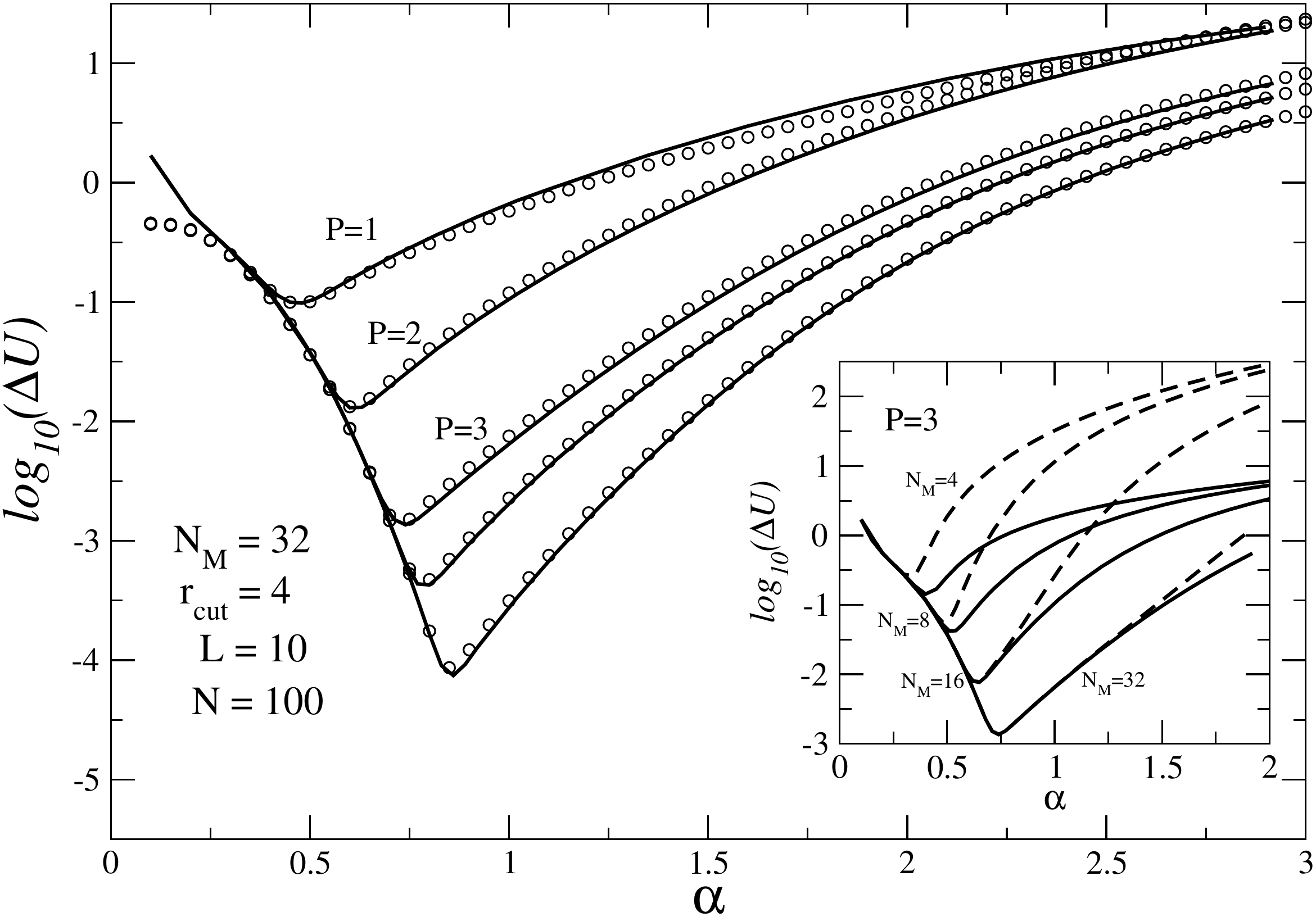}
  \caption{Figure 3}
    \label{f3} 
\end{figure}

\begin{figure}
  \centering
  \includegraphics[width=\linewidth,clip=true]{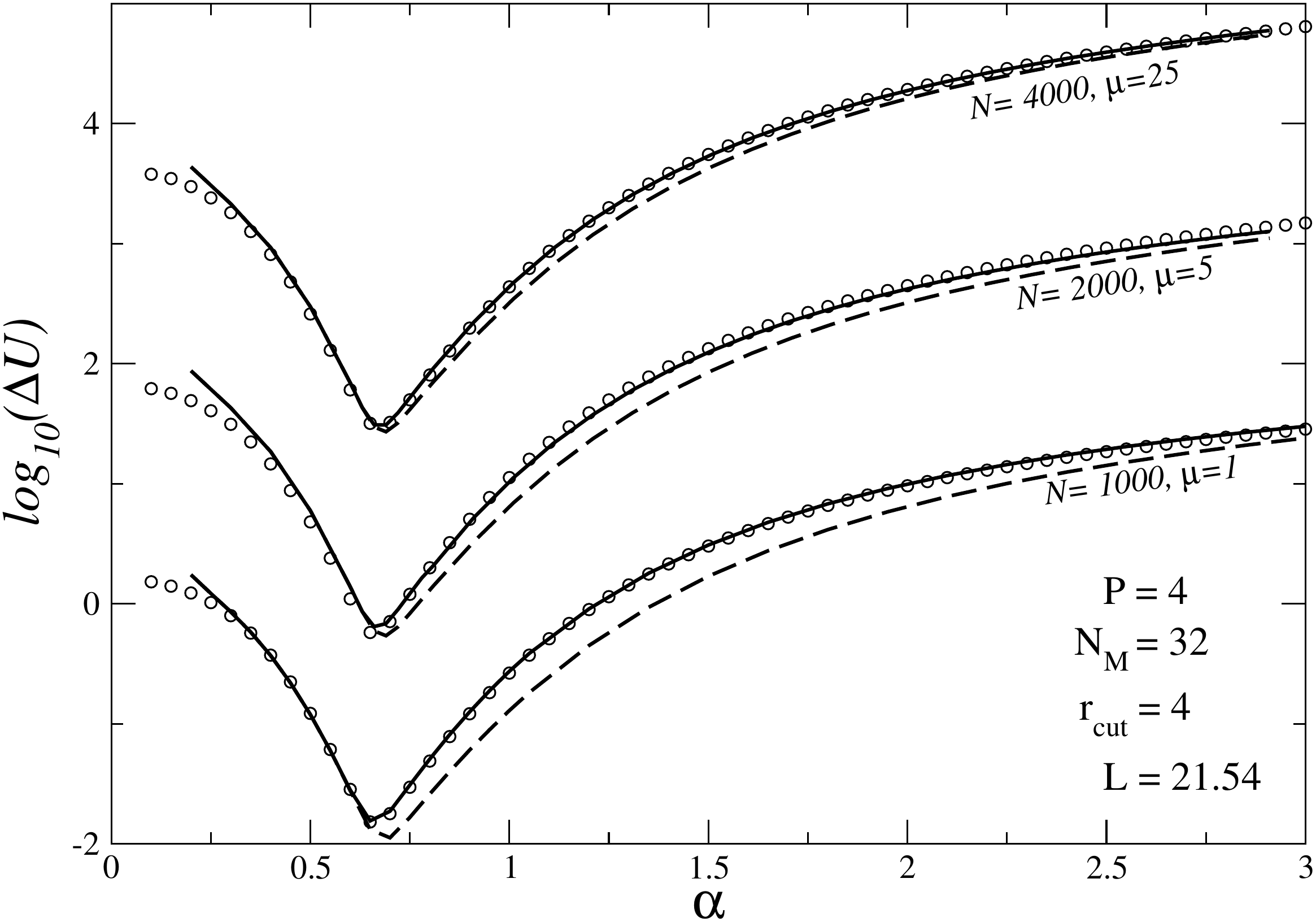}
  \caption{Figure 4}
    \label{f4} 
\end{figure}

\begin{figure}
  \centering
  \includegraphics[width=\linewidth,clip=true]{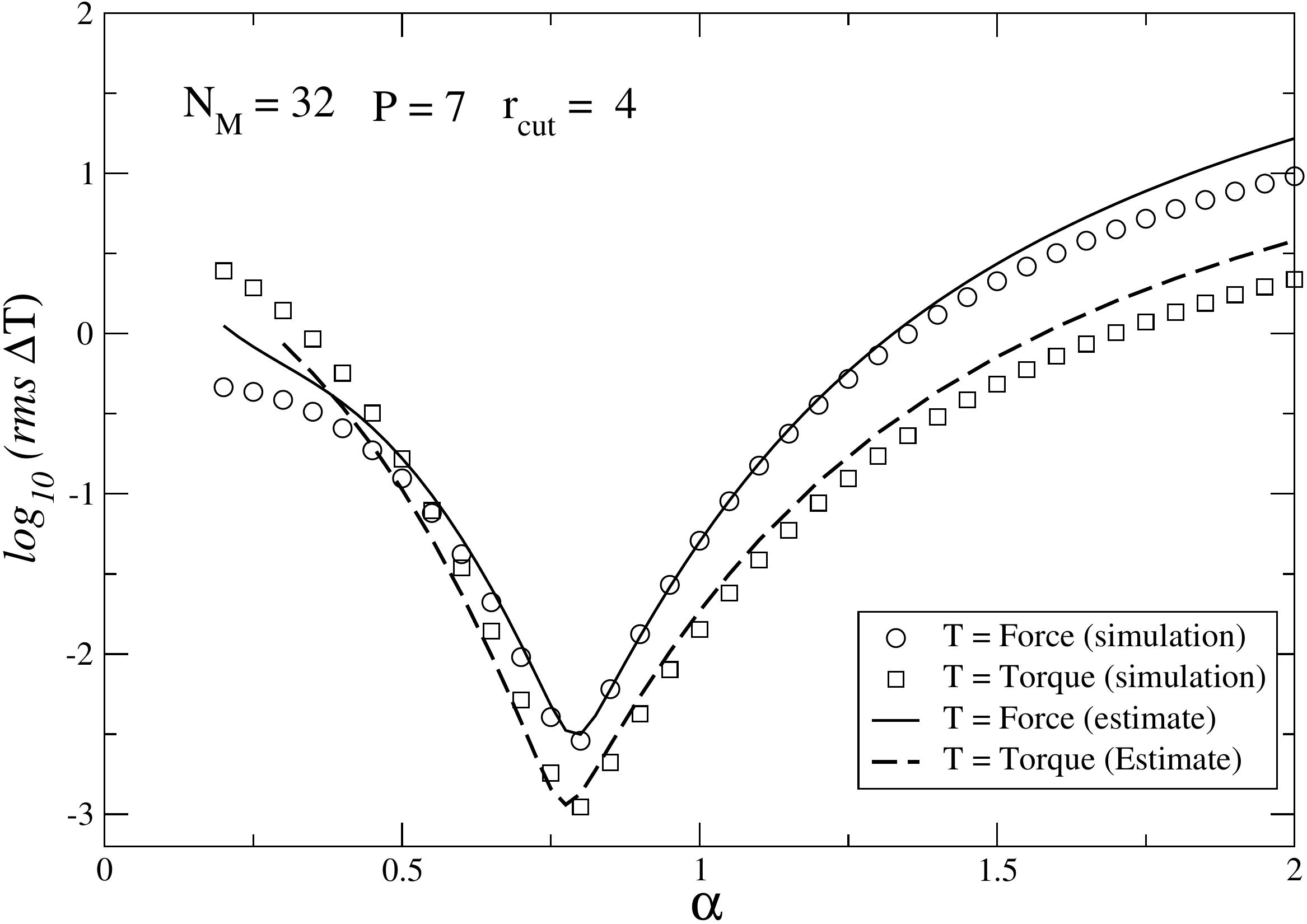}
  \caption{Figure 5}
   \label{f5} 
\end{figure}

\begin{figure}[htbp]
  \centering
  \includegraphics[width=\linewidth,clip=true]{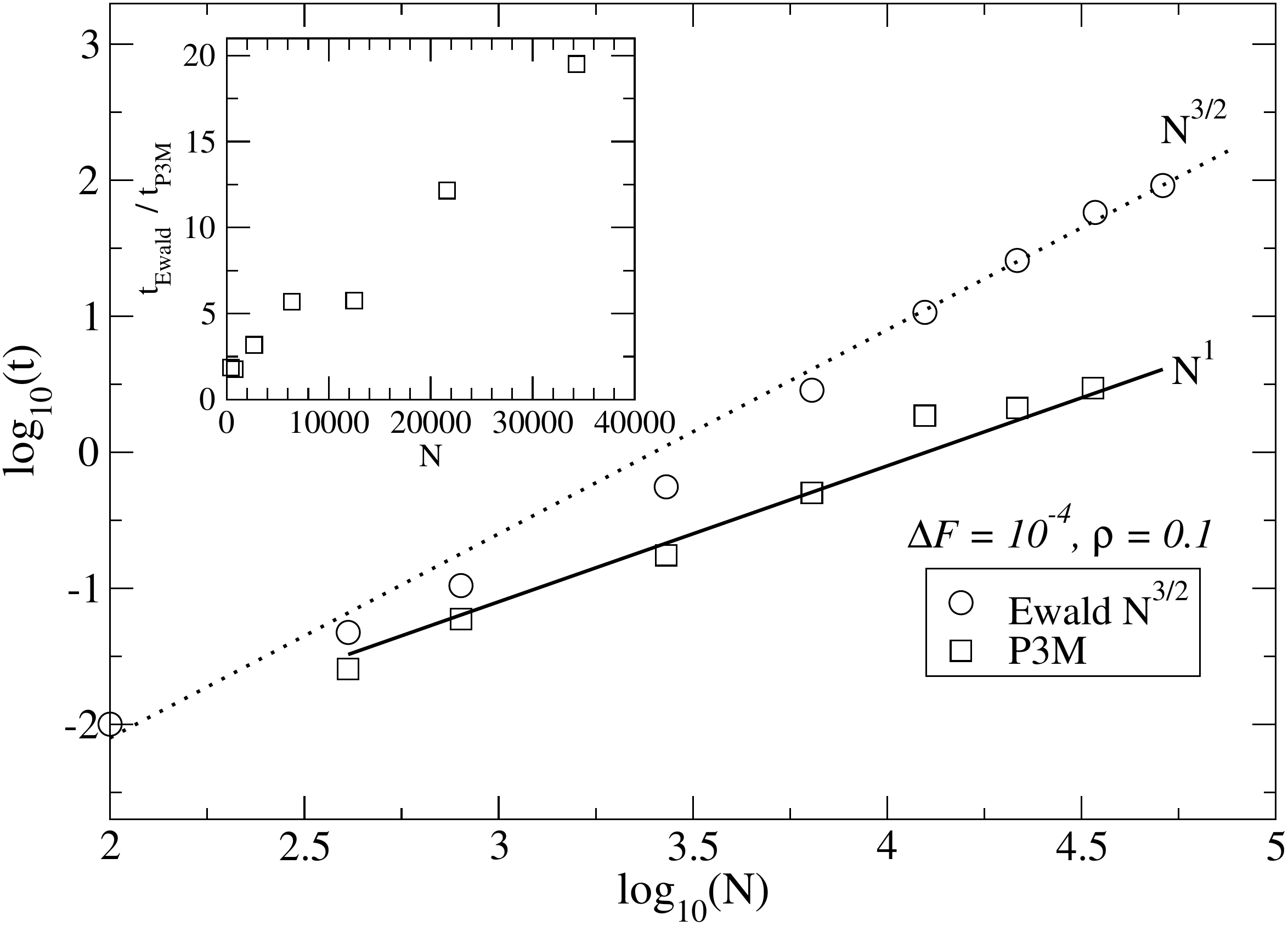}
  \caption{Figure 6}
  \label{f6} 
\end{figure}

\begin{figure}[htbp]
  \centering
  \includegraphics[width=\linewidth,clip=true]{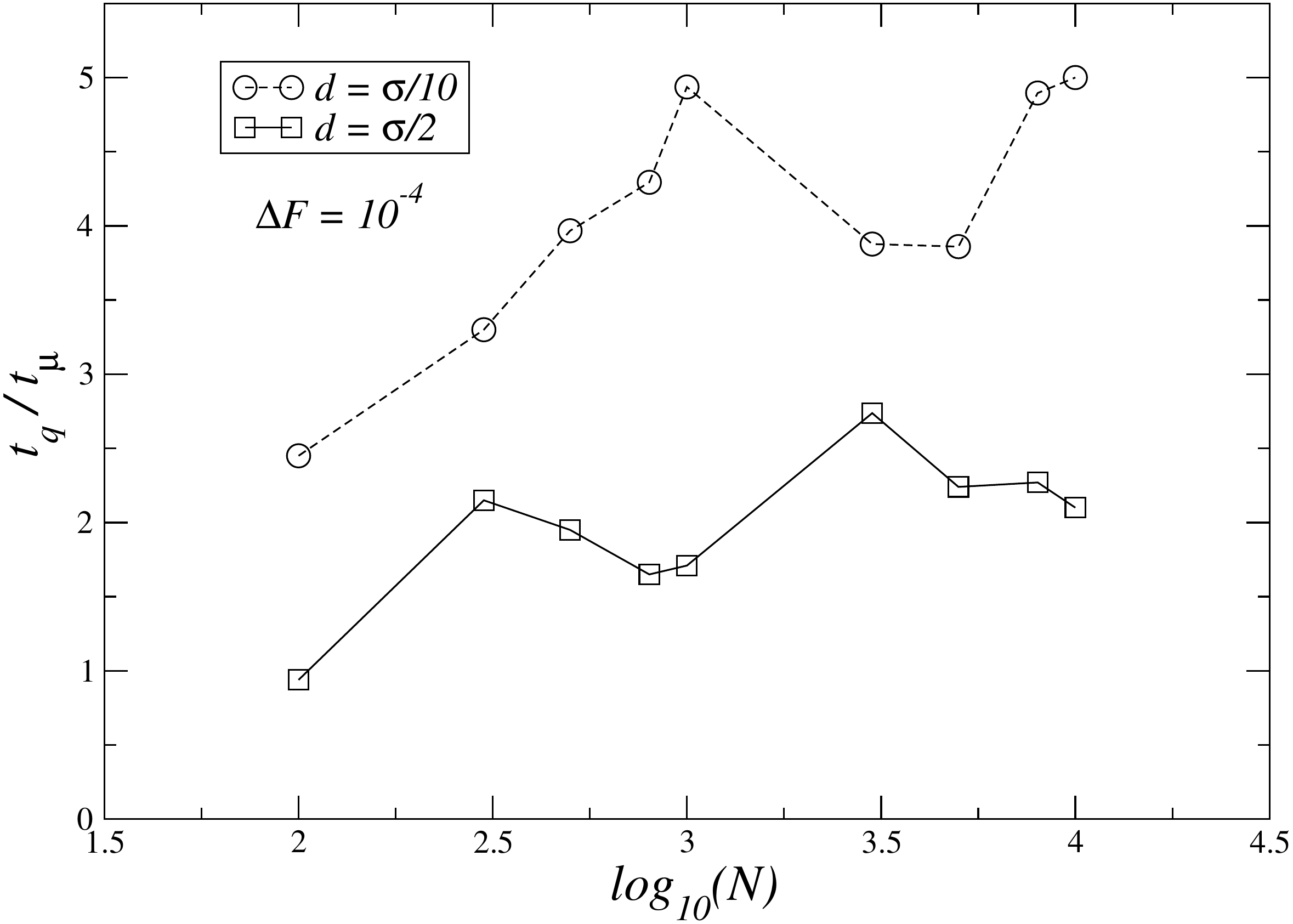}
  \caption{Figure 7}
    \label{f7} 
\end{figure}

\begin{itemize}

\item TABLE I:     Definitions of the various transforms between real-space and reciprocal
    space: Fourier transform of a non periodic function (first line); Fourier
    series of a periodic function (second line); and Finite Fourier transform
    of a mesh-based function (third line).  
    
\item TABLE II:      Exact versus \PPPM\ Madelung-Self interactions. The mean and rms
  error of \madelungself\ interactions are computed by taking an
  average over all positions and orientations of the dipole moment.

\item FIGURE 1:  The rms error $\Delta F$ of the forces (circles) for a
    system of $100$ randomly distributed dipoles with $N_M=32$ mesh
    points and real space cutoff $\rcut=4$. Box size $L=10$. From top
    to bottom, the order of the charge assignment function, $P$, is
    increased from 1 to 7. The solid lines are the theoretical
    estimates (eq.~\eqref{summarydF}).  In the inset, the order of the
    assignment function is $P=3$, and the number of mesh points per
    direction is varied (from top to bottom): $N_M \in \{4, 8, 16, 32,
    64 \}$.  
    
 \item FIGURE 2:    Computational and theoretical rms error of the torques
    $\Delta \tau$ for the same system as in figure \ref{f1}. The
    dotted lines in the inset plot show two examples of the deviations
    observed at large values of the splitting parameter $\alpha$ when
    the errors due to \madelungself\ torques are neglected in the
    evaluation of the rms error estimates (eq.~\eqref{summarydTau}).
 
 \item FIGURE 3: Comparison of the theoretical estimates for the rms errors
    of the energy (eq.~\eqref{summarydU}) (solid line) with the
    corresponding numerical rms errors (circles). Several values of
    the charge assignment order $P \in [1,5]$ are depicted for systems
    with box length $L=10$, number of dipoles $N=100$, cutoff
    parameter $\rcut=4$, and mesh size $N_M=32$.  The numerical rms
    error of the energy is computed averaging over 100 random
    conformations, using eq. \eqref{uest1}.  The inset shows a
    comparison between the rms error obtained using the energy
    correction $U^{(corr)}$ [eq.~\eqref{Ukws}] (solid lines), and the
    rms error when no energy correction is applied (dashed lines) for
    $P=3$ and different mesh sizes $N_M \in \{4, 8, 16, 32 \}$. 
    
\item FIGURE 4:    Similar comparison as in figure \ref{f3} for systems with
    different number of particles and dipole moments:
    $(N=1000,|\bmu|=1)$, $(N=2000,|\bmu|=5)$, and
    $(N=4000,|\bmu|=25)$.  The box length is set to $L=21.54$,
    assignment order $P=4$, and mesh size $N_M=32$.  Dashed lines
    depict the rms errors when \madelungself\ and energy correction
    terms are dropped out from expression \eqref{summarydU}, see
    eq.~\eqref{Uapproximacio}.
    
\item FIGURE 5:    Comparison of the theoretical rms estimates of forces and
    torques predicted for random conformations versus the numerical
    rms errors for a typical conformation in a simulation of a
    ferrofluid system \cite{wang02a}. Number of particles $N=1000$,
    diameter $\sigma=1.58$, dipolar coupling parameter $\lambda =3$,
    and volume fraction $\phi_v = 0.3$. The particles are under the
    influence of an external magnetic field along the z axis
    characterized by a Langevin parameter $\alpha_L = 2.0$. 

\item FIGURE 6:    Time required to compute forces and torques as a function
    of the number of particles in the system using a typical desktop
    computer. The computing time $t$ is given in seconds. Circles
    denote the optimal dipolar-Ewald method, and squares the new
    dipolar \PPPM\ method. In both cases, their respective parameters
    have been tuned to give maximum speed at fixed force-accuracy.
    The accuracy is set to $\Delta F = 10^{-4}$. The density of
    particles is $\rho=N/V=0.1$. Lines with slopes $1$ and $3/2$ are
    plotted to guide the eye. The inset plot shows the relative speed
    of dipolar-\PPPM\ method compared to the fastest dipolar Ewald-sum
    as a function of the number of particles in the system.

\item FIGURE 7:    Relative speed of the dipole-based model to the charge-based model 
 as a function of the logarithm of the number of particles in the system
 (see details for the models in text, Sct. \ref{compamodels}) .
 $t_{\mu}$ and $t_q$ are  the times needed by the dipole-based and the
 charge-based models  respectively to integrate $20000$ time steps.  In
 all  systems the number density is $N/V=0.1$, and the algorithm 
 parameters has been set for each system to the optimal values to 
 yield maximum speed at fixed force accuracy $\Delta F = 10^{-4}$. 
     
\end{itemize}

\clearpage



\begin{table} \label{TableA}
\begin{tabular}{|c|c|c|}
\hline
Period & \textbf{Transform to real space} & Domain \\
\hline
none &
$\displaystyle f(\br) = \frac{1}{(2\pi)^3}\int_{\mathbb{R}^3} \FTnp{f}(\bk) e^{i\bk\dotprod\br} \dd\bk$
&
$\br \in \mathbb{R}^3$
\rule{0mm}{0.5cm}\\[2mm]
$L$ & 
$\displaystyle f(\br) = \mathrm{\FSpletters}^{-1}[\FSp{f}] = \frac{1}{L^3}\sum_{\bk\in\mathbb{K}^3} \FSp{f}(\bk)e^{i\bk\dotprod\br}$
&
$\br \in V$
\\[2mm]
$L$ & 
$\displaystyle f_M(\br_m) = \mathrm{FFT}^{-1}[\FFTs{f}_M]  =
\frac{1}{L^3}\sum_{\bk\in\FFTs{\mathbb{M}}^3}
\FFTs{f}_M(\bk)e^{i\bk\dotprod\br_m}$
&
$\br \in \mathbb{M}^3$\\
\hline
\end{tabular}
\begin{tabular}{|c|c|c|}
\hline
Period & \textbf{Transform to reciprocal space} & Domain\\
\hline
 none 
 &
$\displaystyle\FTnp{f}(\bk) = \int_{\mathbb{R}^3} f(\br)e^{-i\bk\dotprod\br} \dd\br$ &
$\bk\in\mathbb{R}^3$
\rule{0mm}{0.5cm}\\[2mm]
 none 
&
$\displaystyle\FSp{f}(\bk) = \mathrm{\FSpletters}[f] = \int_{L^3}f(\br)e^{-i\bk\dotprod\br} \dd\br$
&
$\bk\in\mathbb{K}^3$ 
\\[2mm]
 $\displaystyle\frac{2\pi}{h}$
&
$\displaystyle\FFTs{f}_M(\bk) = \mathrm{FFT}[f_M] =  h^3\sum_{\br_m\in\mathbb{M}^3}f_M(\br_m)e^{-i\bk\dotprod\br_m}$
&
$\bk\in\FFTs{\mathbb{M}}^3$\\
\hline
\end{tabular}
 \caption{Table I}
\end{table}
\clearpage


\begin{table}[h]\label{TableB}
\begin{tabular}{p{0mm}|c|c|c|c|}
  \cline{3-5}
  \multicolumn{2}{c|}{} & Energy		& Force	& Torque \rule{0mm}{1.05em}\\
  \cline{2-5}
  \multirow{4}{0mm}{}
  & Exact Madelung-Self interaction \rule{0mm}{1.em} & $\displaystyle\frac{2 \alpha^3}{3\sqrt{\pi}}
  - \frac{2\pi}{3 L^3}$\rule{0mm}{1.6em}	& 0	& 0 \\
  & \PPPM\  Madelung-Self interaction \rule{0mm}{1.em} &  eq. \eqref{energymadself}	& 0	& eq. \eqref{taumadself} \\
  & Average error	& eq.~\eqref{average_error}	& 0	& 0	\\
  & Rms error   &eq.~\eqref{Uself2}		& 0	& eq.~\eqref{dtw2}\\
  \cline{2-5}
\end{tabular}
\caption{Table II} 
\end{table}

\clearpage


\clearpage
 
\clearpage

\clearpage

\clearpage


\clearpage


\clearpage


\end{document}